\documentclass[aps,pre,twocolumn,superscriptaddress,,amsmath,amssymb,]{revtex4-1} 
\usepackage{graphicx}
\usepackage{xcolor}
\usepackage{dcolumn}
\usepackage{bm}

\usepackage[utf8]{inputenc}
\usepackage[T1]{fontenc}
\usepackage{mathptmx}
\usepackage{etoolbox}
\usepackage{soul}
\usepackage{diagbox}

\begin{document}


\title{Bulk and  surface dominated phenomena and the formation of  pentagonal structures in 2-D  strongly coupled  finite dust  clusters}

\author{Mamta Yadav}
\email {ymamta358@gmail.com} 
\affiliation{Department of Physics, Indian Institute of Technology Delhi, Hauz Khas, New Delhi 110016, India}

\author{Aman Singh Katariya}
\affiliation{Department of Physics, Indian Institute of Technology Delhi, Hauz Khas, New Delhi 110016, India}

\author{Animesh Sharma}
\affiliation{Department of Physics, Indian Institute of Technology Delhi, Hauz Khas, New Delhi 110016, India}

\author{Amita Das}
\email {amita@iitd.ac.in}
\affiliation{Department of Physics, Indian Institute of Technology Delhi, Hauz Khas, New Delhi 110016, India}

\begin{abstract}This paper explores the prevalence of size-dependent aspects in the context of dust clusters with the help of Molecular Dynamics (MD) simulations in two dimensions. The transition from macroscale (identified by the dominance of the number of dust particles in bulk) to microscale (where the number of particles on the surface dominates) is explored systematically. 
The dust particles organize in a multi-ringed structure under transverse confinement.  The ring size and the number of rings increase with increasing number  of dust particles.   Interestingly,  the formation of an additional ring is always preceded by structures with a pentagonal symmetry in the core.  A detailed study of this formation has been investigated under various symmetries of the boundary condition and different values of the shielding potential.

\end{abstract}
\maketitle

 \section{ Introduction}
  Parallels have often been drawn between the dusty plasma medium and other disciplines such as condensed matter, complex fluids, etc. \cite{morfill2009complex}  For instance, phenomena of phase transition, the appearance of crystalline forms, and their melting under appropriate conditions are observed both in condensed matter and strongly coupled dusty plasmas \cite{hartmann2007molecular,thomas1994plasma}. Furthermore, the dusty plasma also mimics traits of soft matter and complex visco-elastic fluids. It is, therefore, of interest to look for other kinds of relatable aspects. The latest research interest in condensed matter pertains to surface-dominated effects in a finite-sized matter, displaying novel characteristics that are considerably distinct from the bulk properties of the medium. We explore this feature in the context of finite-sized dust clusters.  
  
Dusty (or complex) plasmas are a collection of nano to micrometer-sized solid dust particles suspended in a plasma medium. Usually, the dust particles get charged in the plasma environment by the constant bombardment of plasma electrons and ion species on their surface. When the number of dust particles is significantly large, they start exhibiting collective properties. The plasma can then be considered as a three-component medium in which, along with lighter electrons and ions, the third component is the dust species, which are much heavier ($10^{-18}-10^{-12}$ Kg) and highly charged ($\sim 10^4 e$). The collection of these dust particles, therefore, exhibits collective phenomena at much longer lengths and time scales (compared to the regular electron-ion plasma) and can be tracked with ease by simple diagnostic tools such as a camera. The requirement of sophisticated diagnostics does not arise. Due to the very high charge ($\sim 10^4$ electrons on micrometer-sized dust) acquired by the dust particles, the dust species become strongly coupled. This implies that the Coulomb coupling parameter ($\Gamma$), defined as the ratio of the interparticle potential energy to the kinetic energy of the species, i.e., $\Gamma = Q^2/K_bTa$ ($Q$, $T$, and $a$ are the charge, temperature, and inter-grain distance of species, respectively), can easily exceed unity at even room temperature and normal densities.  So dusty plasma becomes a novel medium to test the properties of a strongly coupled system under normal conditions. The exotic low temperature and superdense regimes,  where the strongly coupled behavior usually manifests for the usual electron-ion plasma can be easily avoided. This system exhibits  crystallization {\cite{hayashi1994observation,thomas1994plasma,chu1994direct}, visco-elastic behaviour {\cite{feng2010viscoelasticity,feng2012frequency,singh2014visco}, and exotic transport properties {\cite{nosenko2008heat,nunomura2005heat,liu2008superdiffusion}. Various well-known fluid instabilities in strongly coupled regimes have been explored with the help of dusty plasmas \cite{merlino1998laboratory,tiwari2012kelvin,das2014suppression,dharodi2022kelvin,veeresha2005rayleigh}. Linear and non-linear waves \cite{kumar2018spiral,das2014exact,kumar2017observation}, collective modes \cite{kaw1998low,kaw2001collective,ivlev2000anisotropic}, phase transitions \cite{thomas1996melting,melzer1996experimental,melzer1996structure,maity2019molecular,schweigert1998plasma} phenomena as well as single particle features  \cite{teng2009wave,maity2018interplay} have been investigated. The dynamical equilibrium of charged dust clusters with Yukawa interaction  \cite{maity2020dynamical} and chaotic evolution  \cite{deshwal2022chaotic} have been shown lately.  
  
Decreasing the temperature of the species can also lead to a high value of the parameter $\Gamma$.   This is achieved in the context of an ultra-cold neutral plasma regime. Various properties of the ultracold neutral plasma have been investigated, i.e., expansion dynamics \cite{robicheaux2002simulation,pohl2004kinetic,laha2007experimental}, formation of Rydberg atoms \cite{killian2001formation}, etc. In MD simulations formation of classical bound structures and the properties of crystalline structures have been studied  \cite{yadav2023structure,YADAV2024134326}. Apart from high charge and low temperatures, the strongly coupled regime can also be achieved by reducing interparticle spacing  $a$, this happens in dense plasmas such as in the inner region of planets and superdense stars \cite{van1991dense}.

In this manuscript, two-dimensional molecular dynamics simulations have been carried out to study the organization of dust structures as the particle number is increased. These particles are charged and interact with  Yukawa interaction amongst themselves. The screening is due to the background plasma particles which are treated as massless and would immediately shield the charge on the dust particle. For the two dimensional simulations these particles are confined by the imposition of external fields. 
As the particle number is increased they get  organizes in various rings around a central core. The particles in the outermost ring constitute surface particles and the remaining particles are the bulk particles. The behavior of the cluster is observed to change as a function of the ratio of the number of particles on the surface to that in the bulk. It is also interesting to observe that the addition of each new ring is preceded by a pattern with pentagonal symmetry at the core of the finite cluster.
   
This article has been organized as follows: In Section II, the simulation technique and simulation parameters are discussed briefly. In Section III, we discuss the formation of equilibrium patterns in the three regimes of scales. The dust particles arrange themselves in a multi-ringed structure.  The inherent dynamics associated with these structures are outlined which depends on the ratio of the dust particles in the outermost ring to that of the remaining particles in the bulk. Section IV seeks to understand the phenomena leading to the addition of new rings (shells). It is shown that a pentagonal structure arrangement is observed at the core preceding the addition of any new shell.  Finally, in Section V, we summarize and conclude.

  \label{intro}

\section{ MD Simulation Details}
\label{mdsim}

In this study, we have carried out a two-dimensional (2D) Molecular Dynamics (MD) simulation with the help of the open-source code LAMMPS \cite{plimpton1995fast}. The dynamical evolution of charged dust clusters has been studied in two dimensions. The charged dust particles are initially distributed randomly in the square simulation box. This avoids any pre-correlation amidst the dust species. The charge ($Q_d$) and mass ($M_d$) of the dust species are chosen to be $11940e$ (where $e$ is the charge of an electron) and $6.99\times10^{-13}$ kg, respectively \cite{nosenko2004shear}. The negatively charged dust particles interact with each other through the screened Coulomb potential, called the Yukawa potential \cite{shukla2015introduction}. This choice takes into account the response of the background lighter electron and ion species of the medium, which screen the dust charge very rapidly.  The Yukawa potential is of the form $U(r)= (Q_d/4\pi\epsilon_or)\exp(-r/\lambda_D)$. Where $Q_d$ and $\lambda_D$ are the charges on the dust particles and plasma Debye length, respectively.  A typical screening length for Yukawa pair interaction between dust particles is represented by the plasma Debye length, $\lambda_D$. All the length scales are normalized by the inter-dust grain distance ($a= 2.285\times10^{-3}$m). So, the normalized screening parameter ($\kappa$) that represents the strength of the pair interaction is represented by $\kappa=a/\lambda_D$. In a two-dimensional (2D) system, the length of the simulation box is taken to be $L_x=L_y=12.7943a$ in the $\hat x$ and $\hat y$ directions, respectively. 
 As dust particles are negatively charged, they repel each other through the Yukawa potential. In order to confine dust particles in the x-y plane, external fields/potential are applied. Various shapes of the confining potential have also been experimented with in the simulation. The total force acting on any $l^{th}$ particle at any time is the sum of the Yukawa interactions by all other particles and the external force ($F_E$) 
 \begin{equation}
     \textbf{F}_l = -Q_d \sum_{m=1}^{N_p} \nabla U(r_l,r_m) + F_E
     \label{equation of motion}
 \end{equation}
where $N_p$ is the total number of dust particles. Also, $r_l$ and $r_m$ define the positions of the $l^{th}$ and $m^{th}$ dust species at a particular time. The specific forms of  $F_E$ chosen in our studies have been described later.  To track the dynamics of dust particles, we chose the simulation time step $0.01\omega_{pd}^{-1}$. In our simulation, all the time scales are normalized with the inverse of dust plasma frequency $\omega_{pd}^{-1}$. To get the necessary thermal equilibrium, the Nose-Hoover thermostat (NVT) \cite{nose1984molecular, PhysRevA.31.1695} has been employed. This thermostat has been utilized to reach the desired value of particle kinetic temperature ($T=208 K$). 
 
\begin{figure*}
 \includegraphics[scale=0.34]{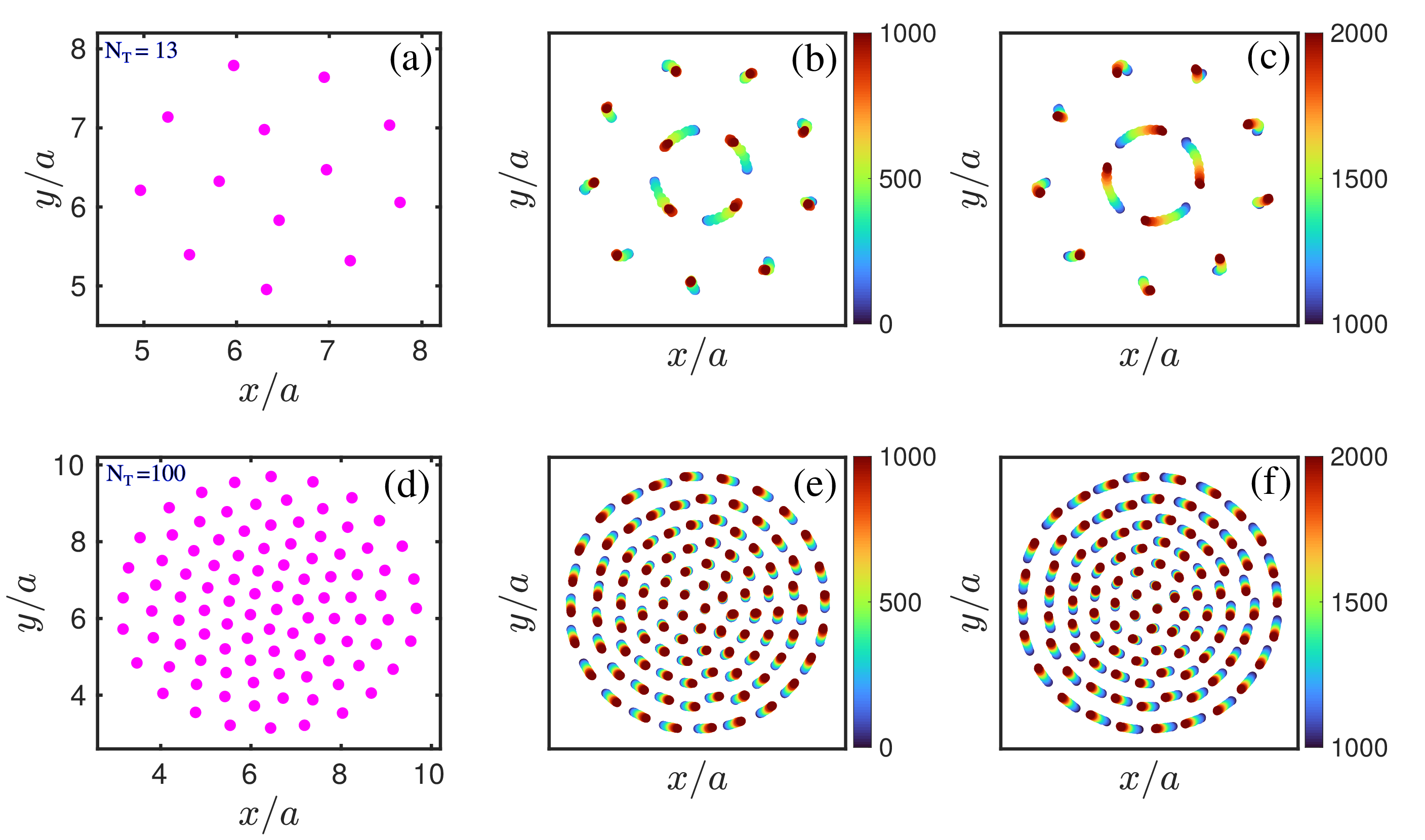}
     \caption{ Subplots (a), (d) represent the arrangments of particles initially for $N_T = 13$ and $100$, respectively. Particle trajectory for $N_T=13$ over the time duration (b) $t\omega_{pd} = 0-1000$ and (c)  $t\omega_{pd} = 1000-2000$. These show the intershell dynamics of the particles. Whereas, particle trajectory for $N_T=100$ over the time duration (e) $t\omega_{pd} = 0-1000$ and (f)  $t\omega_{pd} = 1000-2000$. These show the rigid dynamics of the cluster.  }
  \label{Fig:ref}
 \end{figure*}

 \begin{figure*}
 \includegraphics[scale=0.17]{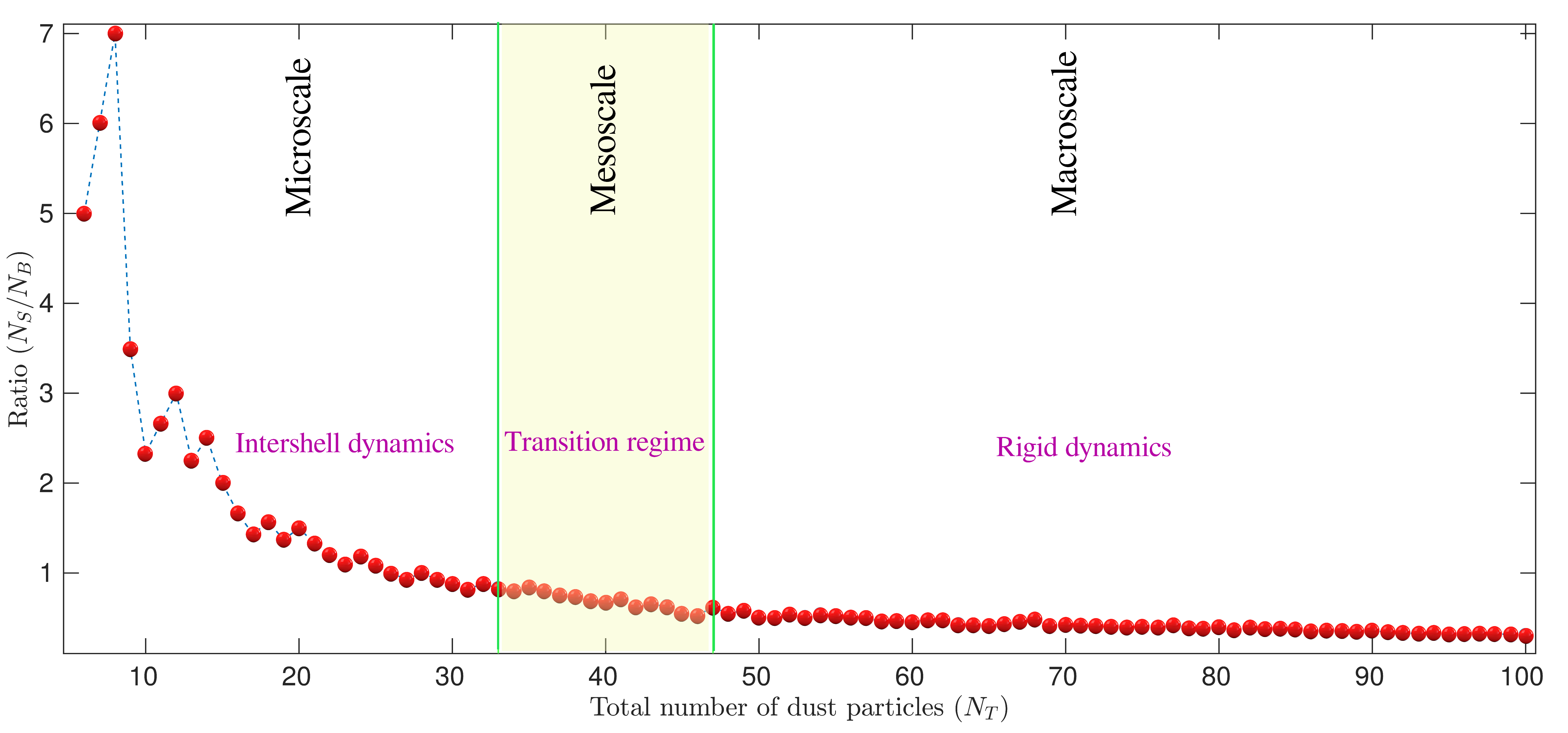}
     \caption{Plot of ratio (number of particles on surface/number of particles in the bulk) to the total number of dust particles. Here, three different regions show the microscale, mesoscale, and macroscale. The mesoscale (represented by yellow shade) shows the transition regime between intershell dynamics and rigid dynamics. The microscale and macroscale correspond to intershell and rigid dynamics, respectively.  }
  \label{Fig:ratio}
 \end{figure*}

 \section {Equilibrium pattern formation}
It is well known that the dust particles strewn inside an electron-ion plasma arrange themselves in a two-dimensional layer, where the gravitational force gets balanced by the electrostatic force at the plasma sheath.  The dust particles are confined in the transverse direction by an applied electric field which forces them to be confined at the center. We have carried out MD simulations to depict such a scenario. We have used the radial electric field of the form  $\textbf{E}(x,y)=K(x-L/2) \hat{x} + K(y-L/2) \hat{y}$  to attain confinement.  The study has been carried out for different numbers of dust particles (from $1$ to $100$) and the equilibrated configurations for each case are ascertained. 

As the number of dust particles in the system is increased they arrange themselves in the form of multiple rings. A new ring appears each time the dust particle number exceeds a certain threshold.  In reference \cite{maity2020dynamical} up to two ringed structures have been studied, starting from $7$ to $15$ number of particles. Some of the structures were found to be static, while others showed a dynamical equilibrium state in which the two rings rotate in opposite directions. This was termed as the intershell rotation. The rotation reversal at intermittent times was observed. However, the direction of rotation remained in the opposite sense for the two rings.  On the other hand, a much higher number of particles (of the order of a few $100$ say)  formed multiple shells, which consistently displayed the behavior of rigid rotation. The ringed shell structure of the cluster and its dynamics have been represented in Fig.\ref{Fig:ref}. Subplots (a), (b), and (c) show the trajectory of $13$ total number of particles ($N_T$), whereas subplots (d), (e), and (f) represent the trajectory of $100$ total number of particles ($N_T$) in the cluster. The initial configuration for $13$ and $100$ number of particles at a particular time is shown by subplots (a) and (d), respectively. The color variation from blue to red shows the time evolution of the cluster. In subplots (b) and (c), the cluster shows the intershell dynamics, while subplots (e) and (f) show the rigid dynamics.   

In the present study, we explore the behavior of transition from inter-shell dynamics to rigid rotation by increasing the number of dust particles one by one from one to hundred. It is observed that the ratio of surface to the bulk number of dust particles plays a strong role in this changed dynamical behavior. We term the particle number in the outermost ring of the cluster as the surface particles $N_s$ whereas the sum of those in the remaining inner rings constitutes the bulk particles $N_B$.  The ratio  $N_s/N_B$ has been plotted as a function of  $N_T$ in Fig.\ref{Fig:ratio}.  The region in $N_T$ corresponding to the inter-shell rotation and rigid oscillations has been depicted in this figure. It clearly shows rigid oscillatory dynamics are observed when the  ratio 
$N_s/N_B < 1$, i.e. when the number of bulk particles dominates. On the other hand, the feature of intershell rotation is observed definitely when the ratio $ N_s/N_T \geq 1$, (i.e., when the surface particle number dominates). However, even for some values of  $ N_s/N_T < 1$ the intershell rotational state is observed provided $N_T < 33$. There is an intermediate region highlighted by a light yellow shading which has been termed as the transition region, in which the state is not very clearly identifiable.

 It is this transition regime that we now discuss. This regime starts at about $N_T \sim 33$ and ends around 
$N_T \sim 47$. For these values of $N_T$ the cluster has a four-ringed structure.
We observe some interesting dynamics when particles start to arrange in the fourth shell. The trajectory of all the particles over the time duration $t\omega_{pd}= 100$ is shown in Fig.\ref{Fig:par_33_37}. Here, the color variation from blue to magenta represents the evolution of particles. In subplot (a), when the innermost shell has only one particle, i.e., $N_p=33$ is shown,  the intershell rotation of particles is observed. The inner particles remain fixed at the center whereas the second and third shell particles move in a clockwise direction, and the fourth shell particles move in an anticlockwise direction. Whereas, when the innermost shell has two particles, the trajectory for $N_p=37$ is shown in subplot (b). This arrangement of particles shows a stationary structure. The particles remain fixed at their positions, displaying merely a small thermal fluctuation. 
 
 \begin{figure}
 \includegraphics[scale=0.3]{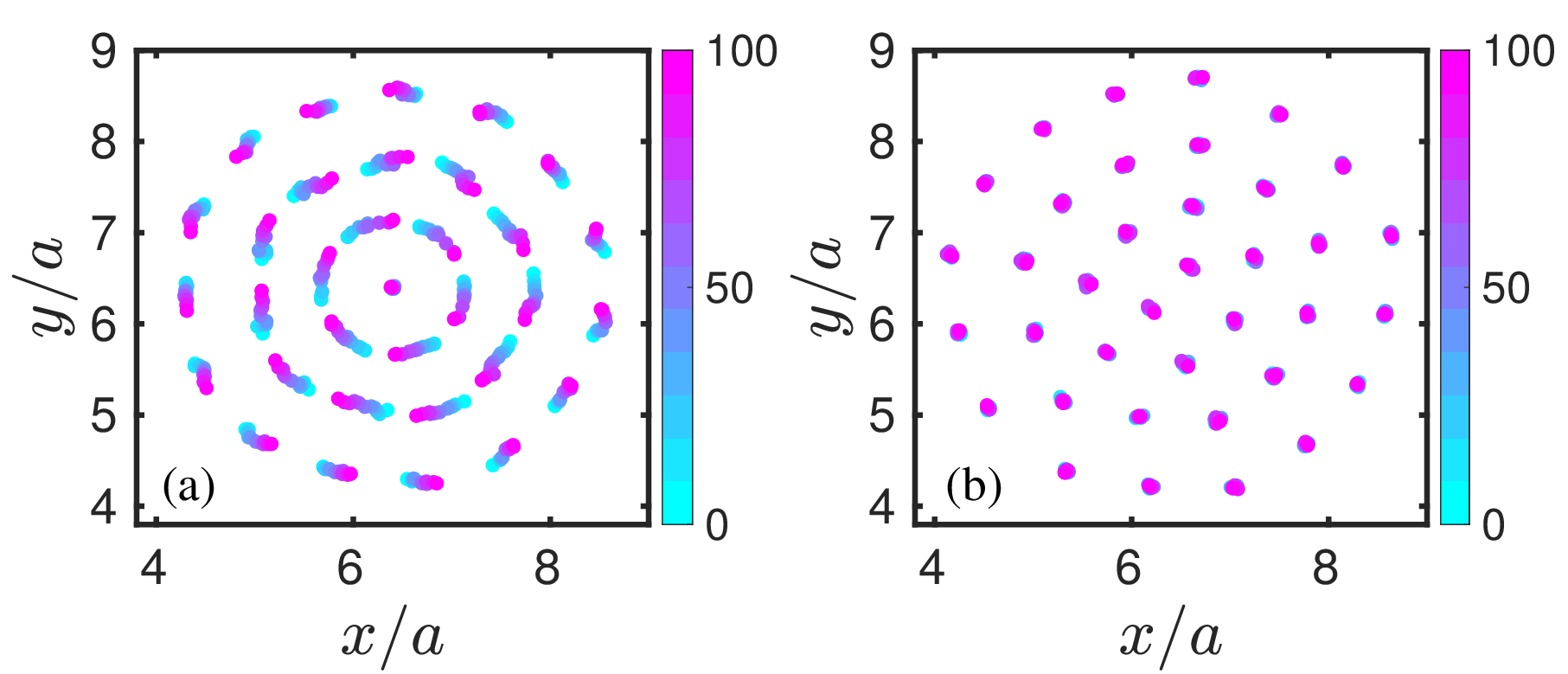}
     \caption{Trajectory of all the particles over time duration $t\omega_{pd}=100$. Subplots (a) and (b) correspond to particles $33$ and $37$, respectively. Here, the color from blue to magenta shows the time evolution of particles.}
  \label{Fig:par_33_37}
 \end{figure}
 
When we further increase the particles,  three particles constitute the innermost shell. Subplot (a) and (b) of Fig.\ref{Fig:par_41_42} represent the time evolution when $N_p=41$ and $N_p=42$, respectively. In subplot (a), when $N_p=41$, for the innermost shell, particles move in the anticlockwise direction. For the particles in the third and fourth shells, their individual dynamics dominate.  Some of them seem to move in clockwise directions others are almost static. They even show some amount of radial fluctuations. Thus one may consider that the particles in the second and third shells are frustrated.
Particles in the fourth shell, however,  appear to remain almost stationary. The other subplot  (b) with a mere addition of one more particle shows a stationary structure where all the particles remain stationary as can be observed from the figure. It should be noted that this happens even though the innermost shell carries the same number of three particles as in subplot (a). 
These configurations are thus, classified by us as in the transition regime where the system exhibits different patterns of evolution.

 \begin{figure}[hbt!]
   \includegraphics[scale=0.27]{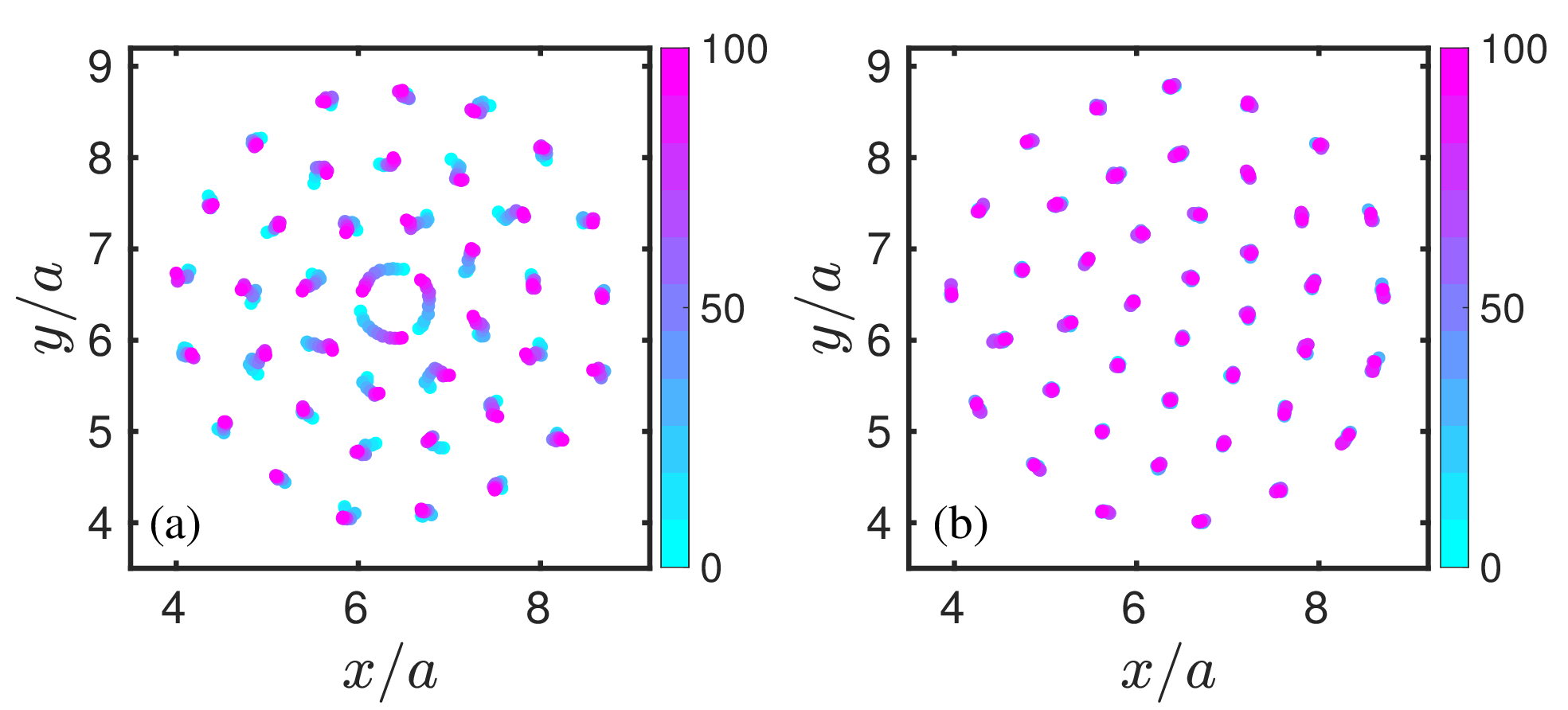}
     \caption{ Trajectory of all the particles over the duration $t\omega_{pd}=100$. Subplots (a) and (b) correspond to $N_p=41, 42$, respectively. Here, the color variation from blue to magenta represents the evolution of particles. } 
  \label{Fig:par_41_42}
 \end{figure}

When we choose $N_T=47$  the innermost shell has four particles. From this onwards with an increasing number of particles,   the state shows rigid oscillations instead of intershell dynamics or following a frustrated scenario of the transition regime. We implement the diagnostics to measure the rigid oscillations through  the average angular displacement $\theta(t)$ defined by 
\begin{equation}
    \theta(t) = \frac{1}{N} \sum_{i=1}^{N} \left[tan^{-1}\left(\frac{Y_i(t)}{X_i(t)}\right)-tan^{-1}\left(\frac{Y_i(t_o)}{X_i(t_o)}\right)\right]
    \label{equation:theta_eqn}
\end{equation}

Here $N = N_T$ the total number of particles, and $t_o$ denotes an arbitrarily chosen initial time.  Furthermore, $X_i(t)$ and $Y_i(t)$ are the coordinates of the location of the particle from the center at any time $t$, i.e.  $X_i= x_i-L/2$ and $Y_i=y_i-L/2$, respectively.
\begin{figure}[hbt!]
 \includegraphics[scale=0.095]{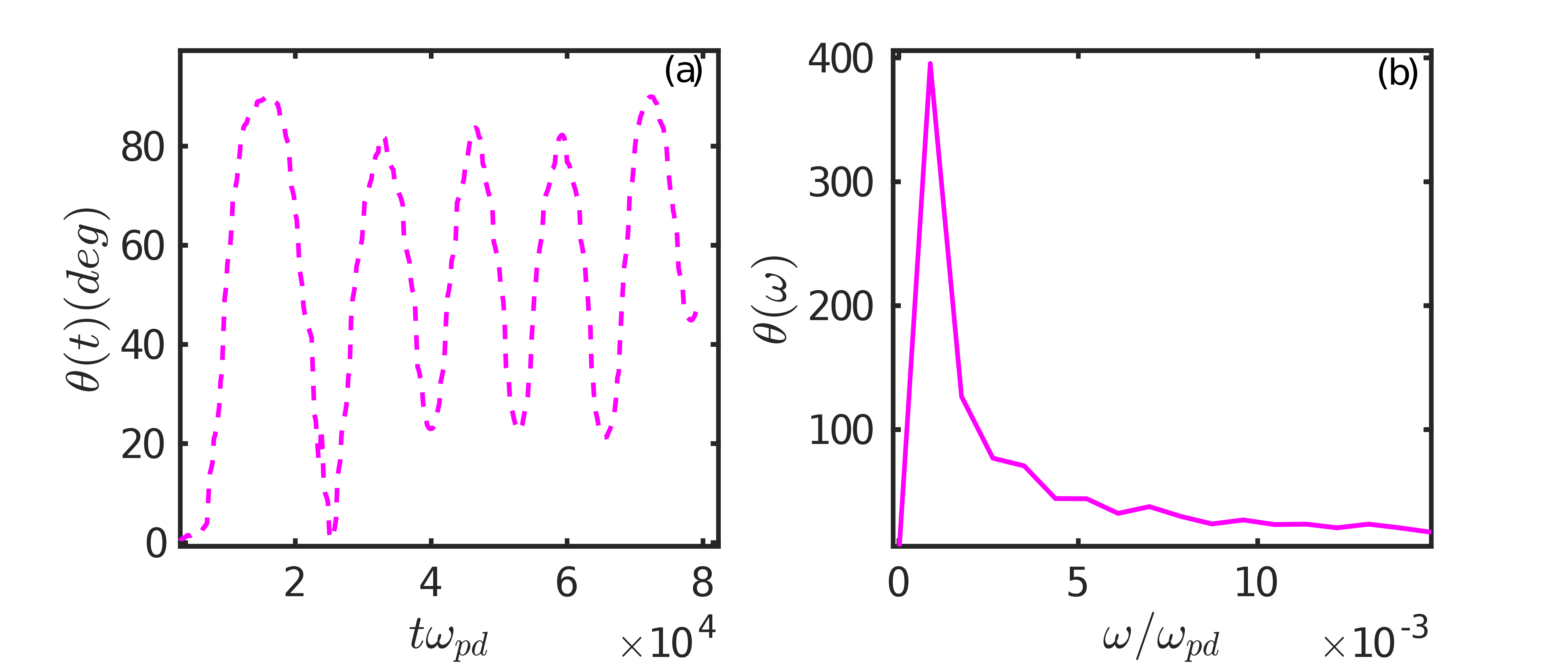}
     \caption{ (a) Time evolution of average angular displacement ($\theta(t)$) for $N_p=47$. (b) The Fourier transform of $\theta(t)$ as a function of normalized frequency $\omega$. }
  \label{Fig:par47}
 \end{figure}

Subplot (a) of Fig.\ref{Fig:par47} represents the time evolution of average angular displacement ($\theta(t)$) for $N_p=47$. The cluster displays rigid oscillation.   The cluster oscillates with a characteristic single frequency as depicted by the Fourier transform of $\theta(t)$ in subplot (b).  The plot shows only one dominant peak, which demonstrates that the cluster oscillation has a single frequency. 
\begin{figure}[hbt!]
 \includegraphics[height = 4.0cm,width = 9cm]{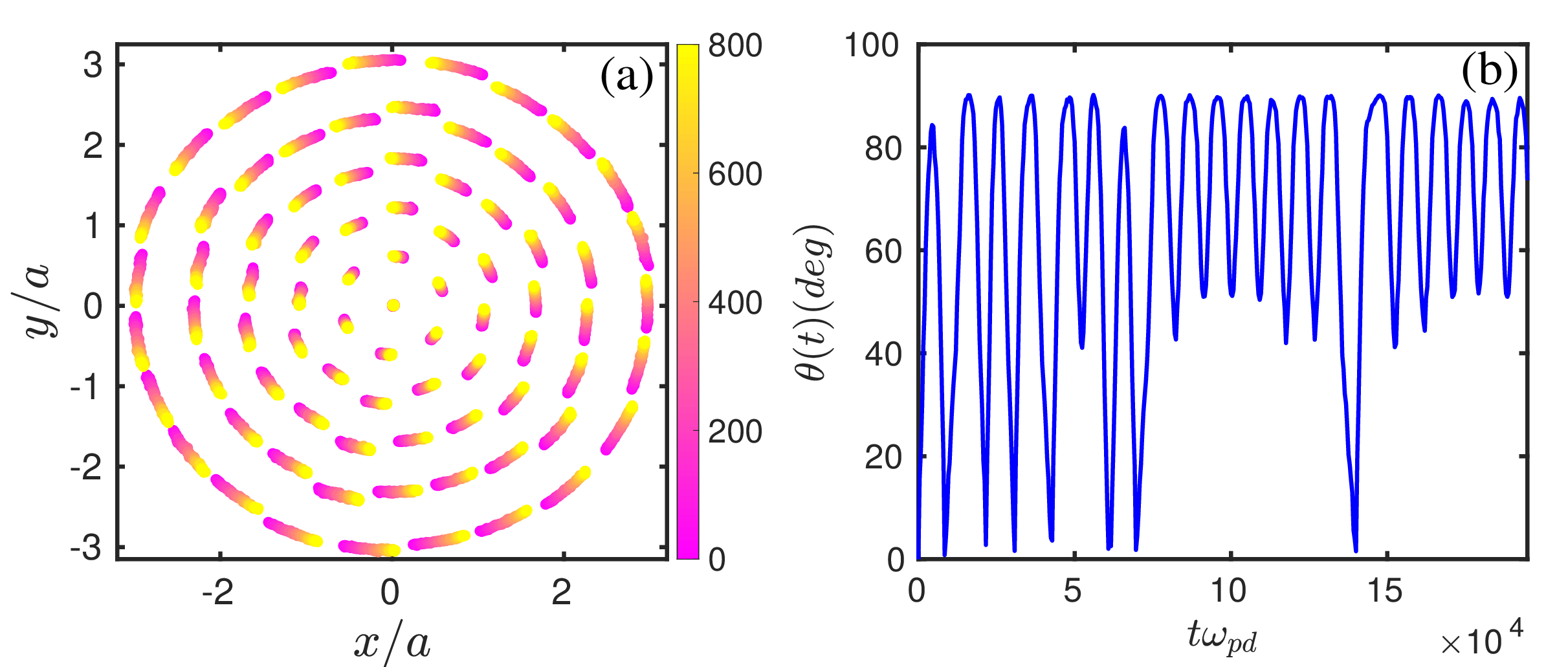}
     \caption{ (a) Particle trajectory for $N_p=80$. Here, the color variation from magenta to yellow shows the evolution of time. (b) Time evolution of average angular displacement ($\theta(t)$) for $N_p=80$.  }
  \label{Fig:80_k1}
 \end{figure}
When we increase the number of particles beyond this, the cluster consistently exhibits rigid oscillations. The subplot (a) of Fig.\ref{Fig:80_k1} represents the trajectory of all the particles up to the time  ($t\omega_{pd}=800$). All the particles show rigid oscillations. To emphasize more, we calculate the average angular displacement ($\theta(t)$) using Eq.\ref{equation:theta_eqn} as shown in subplot (b). From here, we can see that the particles oscillate around a mean value, and the amplitude of the oscillations in this case, is limited to $90^{o}$. This indicates that the rigid oscillations occur, however, there are no complete rotations. 

It is thus clear that the dynamics displayed by the cluster are distinct in the macroscopic and microscopic-sized dust particle clusters, where the bulk and surface effects dominate respectively. The transition between the two occurs through a mesoscopic regime which has been identified clearly by us.

\section{Pentagonal structures}
We now discuss in this section another interesting observation about the formation of structures with pentagonal symmetry in the core of the dust cluster under certain conditions. The green-colored spherical markers in Fig.\ref{Fig:penta_ratio} show the value of $N_T$ when structures with pentagonal symmetry get formed.  These spheres appear for only certain specific values of $N_T$.  Their appearance, however,  is not restricted to the dominance of bulk or surface effects. The value of the ratio $N_S/N_B$ at which they form is not restricted to the domains of macro, meso, or micro regimes.  They appear together in clusters of two and three in this plot. One also observes that each subsequent appearance of a bunch with pentagonal symmetry occurs after a certain value of $\Delta N_T$ which increases monotonically. A closer look at the data reveals that 
 whenever a new shell gets added (with increasing $N_T$), the pentagonal structures precede the same. For instance, it first appears in a single-ringed structure of (0,5), thereafter  (1,5) forms which is preferred over the structure (0,6). Here the first number inside the bracket denotes the particles at the core and subsequent ones in subsequent rings/shells. The second occurrence happens is of (5,10) particles arranged in pentagonal symmetry, which precedes the (1,5,10) arrangement. The third occurrence is when (5,11,14) arrangement precedes (1,5,11,14) and (1,5,11,15). The fourth occurrence happens when (5,11,16,16) is followed up with (1,5,11,16,16) and (1,5,11,16,17). Similarly, the other subsequent arrangements occur. This has been summarised clearly in Table \ref{tab:radial_conf}.

\begin{figure*}
 \includegraphics[scale=0.35]{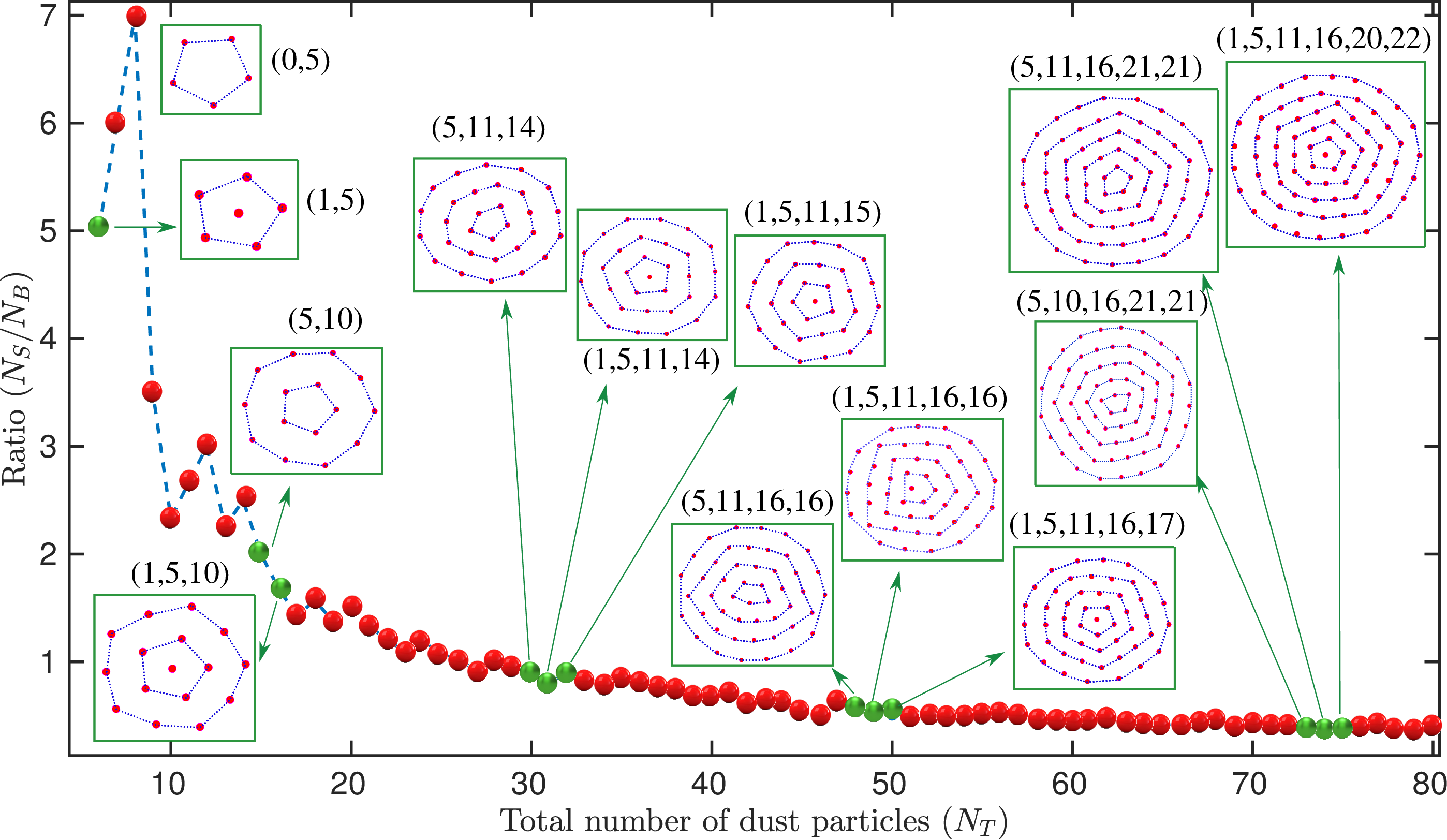}
     \caption{Formation of pentagon structure when a new shell is added. The magenta dots represent the position of particles, and the blue dotted lines show the formation of pentagon structures.  }
  \label{Fig:penta_ratio}
 \end{figure*}

\begin{table}[h!]
    \centering
    \begin{tabular}{|c|c|}
        \hline
No. of Particles & Configuration  \\ 
\hline
 5 & (0,5)  \\
 \hline
 6 & (1,5)  \\ 
 \hline
  15 &  (5,10)       \\
  \hline
  16   & (1,5,10)        \\
  \hline
  30    & (5,11,14)         \\
  \hline
  31      &  (1,5,11,14)       \\
  \hline
  32       &   (1,5,11,15)      \\
  \hline
   48   &   (5,11,16,16)  \\
    \hline
   49   &   (1,5,11,16,16) \\
   \hline
   50     &  (1,5,11,16,17)       \\
   \hline
   73  & (5,10,16,21,21) \\
   \hline
   74  & (5,11,16,21,21) \\
   \hline
   75 & (1,5,11,16,20,22) \\
 \hline
    \end{tabular}
    \caption{Formation of pentagonal structures in radial confinement}
    \label{tab:radial_conf}
\end{table}

The reason for the formation of pentagonal structures in the core of the structure around the time a new shell gets added up has the same basis as that of having a preference for (1,5) in contrast to  (0,6) for a six-particle cluster. It was shown in \cite{deshwal2022chaotic} that the potential energy minimization for 2-D dust clusters under radial confinement occurs for the case of (1,5) configuration. Geometrically also arranging six particles with mutually repulsive interactions under a radially confining potential will be the (1,5) configuration. This is so because for a fixed lattice spacing (interparticle distance between particles determined by the mutual interaction potential), the (1,5) arrangement will be confined in a smaller radius than the (0,6) arrangement. The additional particles always get added up from the center outwards. Thus when an additional ring has to emerge it is preceded by a $5$ particle arrangement in the innermost shell.  Thereafter, (1,5) structure emerges in the core, and then either (1,6) forms, or if the cluster size is large then particles start adding up in higher shells. It is for this reason that the structures with pentagonal symmetry appear in pairs or triplets around the value of $ N_T $ where an additional shell appears.

\subsection{Role of the symmetry of confining potential} 
The external potential defines the typical size of the structure. 
The arrangement for a given number of particles which minimizes the size is preferred for any kind of confining potential profile. A question arises as to whether the pentagonal structures will form if the radial symmetry of the applied potential gets broken up. We have explored other kinds of externally applied potential in our simulations. Simulations with square, pentagonal, and hexagonal boundaries in the 2-D plane were also carried out. The observations have been reported in Fig. \ref{Fig: boundaries}. The green-colored spherical marker represents the formation of pentagonal structures corresponding to that $N_T$, whereas the yellow cross symbol shows the missed pentagonal structures as earlier obtained in radial confinement. It is also summarised in Table \ref{tab:penta_conf_all} for different confinement boundaries. It is clear that in these cases too the formation of a pentagonal structure is always associated with the addition of an extra shell. In some cases (e.g. for boundaries with squares, hexagonal, and pentagonal symmetries), the pentagonal structure formation gets missed even when a new ring adds up.

 \begin{figure*}[hbt!]
   \includegraphics[scale=0.85]{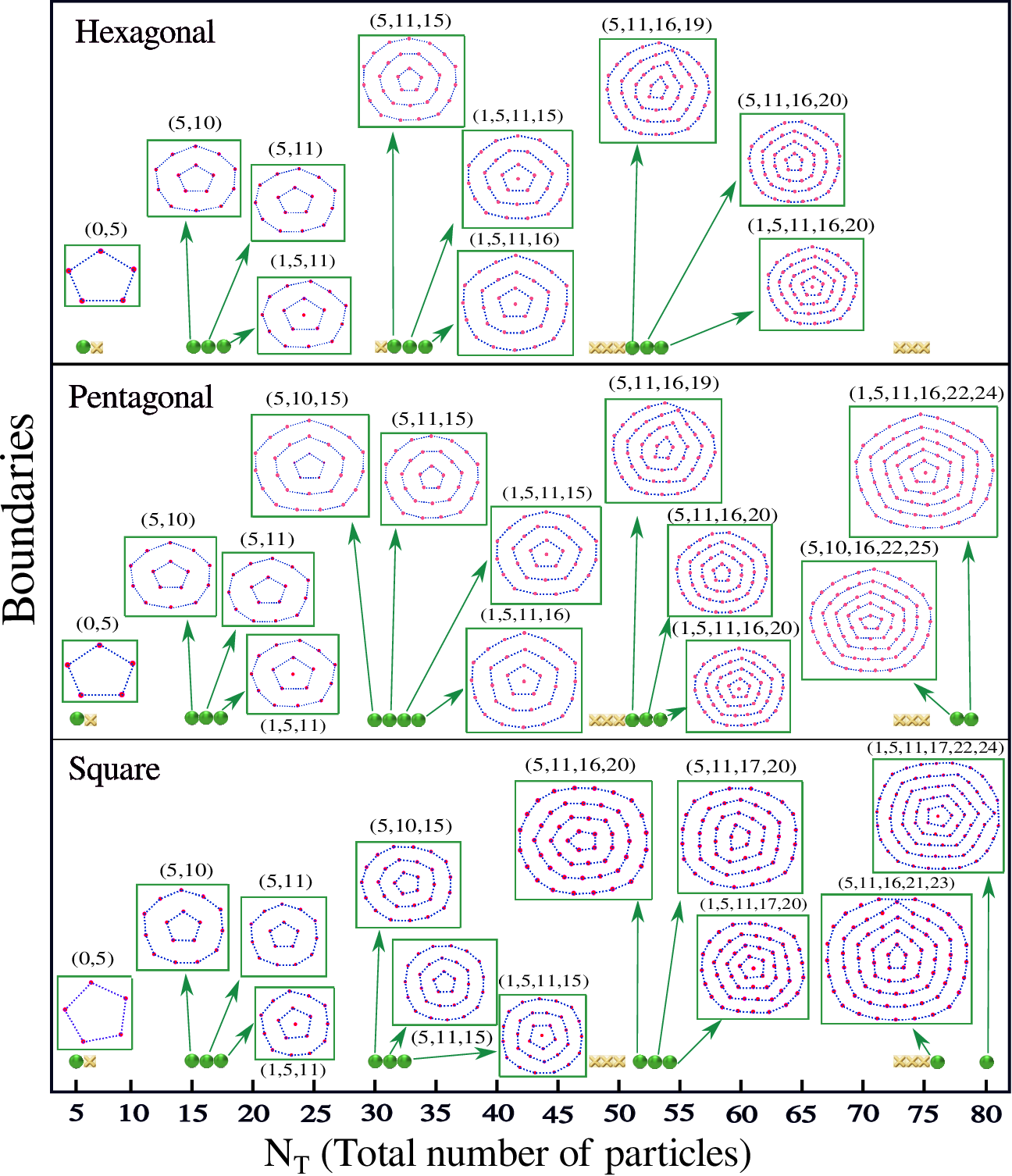}
     \caption{ Formation of pentagonal structures with the total number of particles in different confinement boundaries. The green-colored spherical markers represent the presence of pentagonal structures, and the yellow cross represents the missed pentagonal structures earlier obtained in radial confinements.} 
  \label{Fig: boundaries}
 \end{figure*}

\begin{table*}[t]
\centering
\begin{tabular}{|c|p{8em}|p{8em}|p{8em}|p{8em}|}
\hline
\diagbox[width=10em]{No. of \\particles $\downarrow$}{Confinement \\Boundaries $\rightarrow$}&
Square&Pentagonal&Hexagonal \\ \hline
5 & (0,5) & (0,5) & (0,5)  \\ \hline
15 & (5,10) & (5,10) & (5,10)  \\ \hline
16 & (5,11) & (5,11)& (5,11)   \\ \hline
17 & (1,5,11) & (1,5,11) & (1,5,11) \\ \hline
30 & (5,10,15) & (5,10,15) & -  \\ \hline
31 & (5,11,15) & (5,11,15) & (5,11,15)  \\ \hline
32 & (1,5,11,15) & (1,5,11,15) & (1,5,11,15)  \\ \hline
33 & - &(1,5,11,16) & (1,5,11,16)  \\ \hline
51 & - & (5,11,16,19) & (5,11,15,19) \\ \hline
52 & (5,11,16,20) & (5,11,16,20) & (5,11,16,20)  \\ \hline
53 & (5,11,17,20) \newline (distorted) & (5,11,17,20)& (5,11,17,20)  \\ \hline
54 & (1,5,11,17,20) & - & -  \\ \hline
\end{tabular}
\caption{Configurations of pentagonal structures formed under different confinement boundaries. Here, the screening parameter $\kappa=1$.}
    \label{tab:penta_conf_all}
\end{table*}

\begin{figure*}[hbt!]
    \includegraphics[scale=0.4]{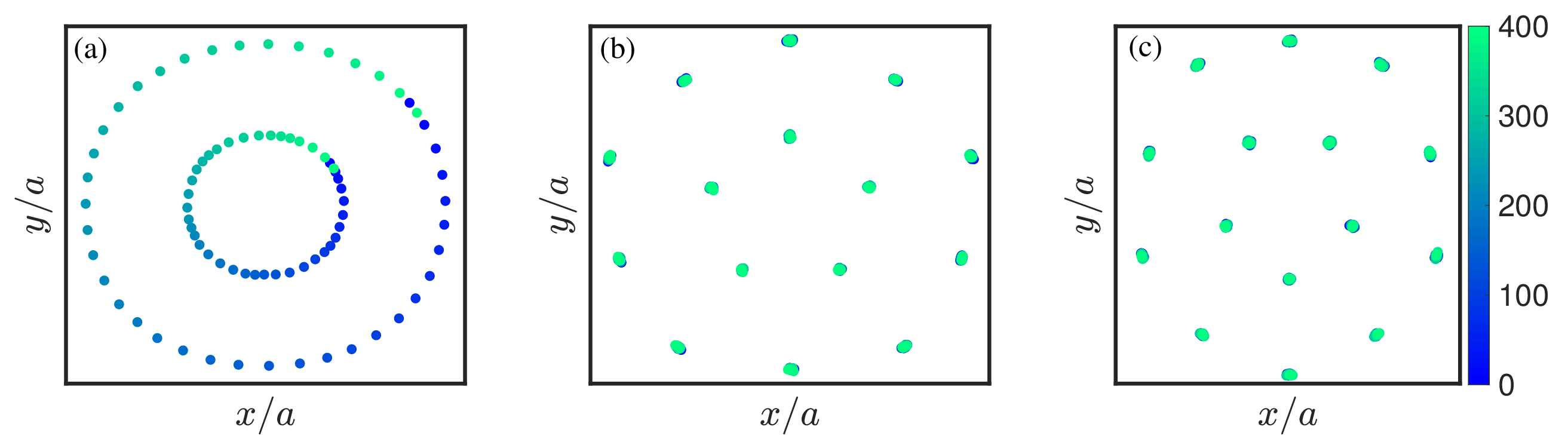}
      \caption{ Trajectory of $N_p=15$ in different confinements. Subplot (a) The trajectory of one particle from each shell in square confinement over the duration $t\omega_{pd}=400$. Subplots (b) and (c) show the trajectory of all particles in the pentagon ($t\omega_{pd}=400$) and hexagon ($t\omega_{pd}=400$) shaped confinements. Here, color variation from blue to green shows the evolution of particles. In this case the screening parameter ($\kappa = 1$) } 
   \label{Fig: track_k1}
  \end{figure*}

 However, the shape of the confinement boundaries affects the dynamics of these dust particles. We consider $N_t=15$, in all confinements, the inner and outer shells contain five and ten dust particles, respectively. The inner shell contains five particles that generate the pentagon structure. Fig.\ref{Fig: track_k1} illustrates the time evolution of particle dynamics in different confinements. Here, the color variation from blue to green shows the particle location changing with  increasing time. Subplot (a) shows the dynamics of one particle from each shell in the square confinement boundary. In both shells, particles rotate in the same direction over this time duration $t\omega_{pd}=400$. Whereas, the pentagon and hexagonal confinement boundaries, the trajectory of all the particles in both shells is shown in subplots (b) and (c), respectively. The particles remain stationary over this time duration $t\omega_{pd}=400$.


\subsection{Role of screening }
We now study the role of the screening parameter which ascertains the interparticle distance in the formation of pentagonal structures. Increasing the value of $\kappa$ results in a reduction in the influence of the interaction potential. It is clear that irrespective of the screening strength the pentagonal structures form again when a new shell gets added to the system. There are, however,  some differences that need to be noted.  

 An increase in the value of $\kappa$ also affects the formation of pentagon structures in different confinements. In Table \ref{tab:penta_conf_all_k2} details of the formation of pentagon structures in square, pentagonal, and hexagonal confinement are shown, respectively. The number of pentagon structures for the hexagonal boundary is almost the same as in the case when screening parameter $\kappa=1$. But there is a great difference in this when the confinement shapes are square and pentagonal, as shown in Table \ref{tab:penta_conf_all_k2}. When the screening parameter increases, the effects of confining force dominates over inter-particle interaction. So shape of confinement plays an important role in this case. In square confinement, we observe only two pentagon structures with $N_p=5,16$; after that, particles are arranged in square shapes. However, in pentagonal confinement, when the number of particles increases, all try to arrange in the pentagon geometry due to the confinement forces dominating. As in Table \ref{tab:penta_conf_all_k2} after $N_p=50$ all the configurations we obtained in the pentagon shape.

\begin{table*}[t]
\centering
\begin{tabular}{|c|p{8em}|p{8em}|p{8em}|p{8em}|}
\hline
\diagbox[width=10em]{No. of \\particles $\downarrow$}{Confinement \\Boundaries $\rightarrow$}&
Square&Pentagonal&Hexagonal\\ \hline
5 & (0,5) & (0,5) & (0,5)  \\ \hline
15 & - & - & (5,10)  \\ \hline
16 & (5,11) & (5,11)& (5,11) \\ \hline
17 & - & (1,5,11) & (1,5,11)  \\ \hline
31 & - & - & (5,10,16)  \\ \hline
32 & - & (5,11,16) & (5,11,16)  \\ \hline
33 & - &(1,5,10,17) & (1,5,11,16) \\ \hline
34 & - &(1,5,11,17) & -  \\ \hline
51 & - & (5,10,15,21) & (5,11,15,19)\\ \hline
52 & - & - & (5,11,16,20)  \\ \hline
53 & - & (1,5,10,15,22) & (5,11,17,20) \\ \hline
54 & - & (5,10,17,22) & (5,11,16,22) \\ \hline
55 & - & (5,10,17,23) & (1,5,11,16,22)  \\ \hline
56 & - & (5,10,18,23) & (1,5,11,17,22)  \\ \hline
57 & - & (1,5,10,18,23) & (1,5,11,17,23)  \\ \hline
58 & - & (1,5,11,18,23) & -  \\ \hline
59 & - & (1,5,12,18,23) & -  \\ \hline
60 & - & (1,5,10,18,24) & -  \\ \hline
\end{tabular}
\caption{Configurations of pentagonal structures formed under different confinement boundaries when $\kappa$=2.}
    \label{tab:penta_conf_all_k2}
\end{table*}

The dynamics of the cluster in the case of different confinement geometries at times are also quite interesting. This has been illustrated in Fig.\ref{Fig: track_k2} for $15$ particles.  
 \begin{figure*}[hbt!]
   \includegraphics[scale=0.32]{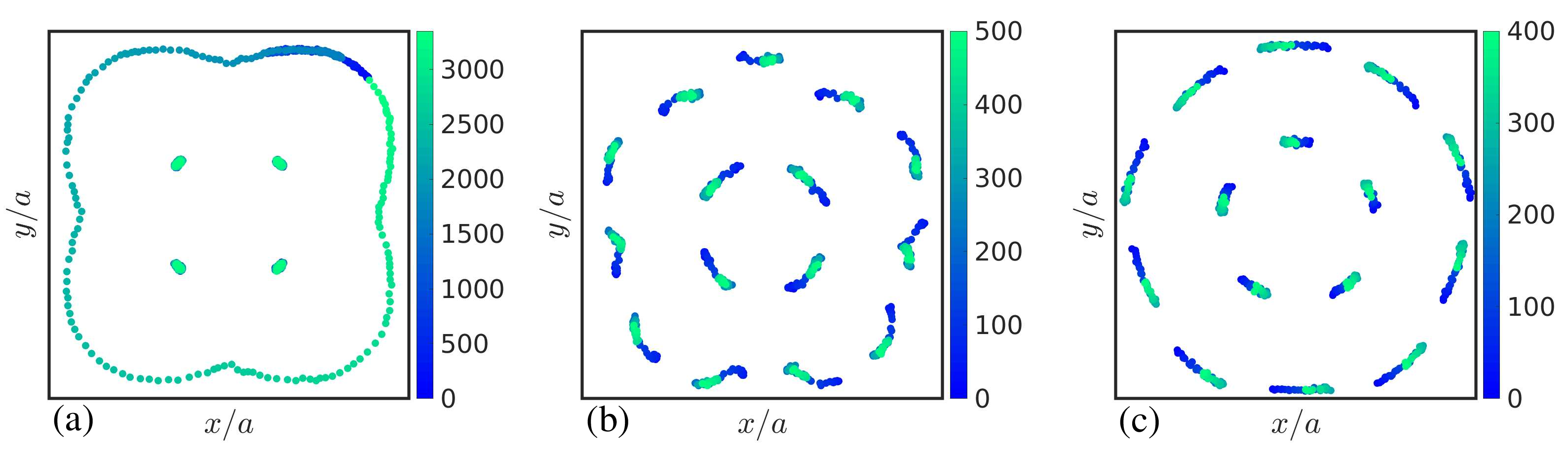}
     \caption{ Trajectory of $N_p=15$ in different confinements. Subplot (a) The trajectory of one particle from the outer shell and all particles from the inner shell in square confinement over the duration $t\omega_{pd}=3350$. Subplots (b) and (c) show the trajectory of all particles in the pentagon ($t\omega_{pd}=500$) and hexagon ($t\omega_{pd}=400$) shaped confinements. Here, color variation from blue to green shows the evolution of particles. In this case, the screening parameter ($\kappa$) is two.} 
  \label{Fig: track_k2}
 \end{figure*}
Subplot (a) shows the trajectory of all the particles from the inner shell and only one particle from the outer shell over the time $t\omega_{pd}=3350$. The particles in the inner shell are almost static and form a perfect square aligning with the boundary symmetry.  In contrast, the particles that make up the outer shell though typically organized in a square geometry are altogether $11$ in number. So the perfect square symmetry is absent.  As a result, the particles suffer unbalanced forces and show evolution. The evolution of one of the particles has been shown in this figure. The trajectory is circular at the four corners of the square and has a cusp at the sides of the square. For pentagonal confinement boundaries intershell dynamics is observed as shown in subplot (b). For hexagonal boundaries (subplot (c)), the trajectories are smooth and shows rigid dynamics.


 \section{Summary and conclusion }

 A two-dimensional dust cluster organization and dynamics have been studied through Molecular Dynamics simulations in the presence of transverse confining forces. The dust particles interact with a screened Coloumb potential, 
 which represents the instantaneous screening by the lighter species of the underlying electron-ion plasma. The dust particles are observed to organize in the form of shells and demonstrate interesting dynamics.  The distinct dynamics of inter-shell rotation and rigid rotation are displayed by micro-sized (when the number of surface particles dominates the bulk number ) and macro-sized (when bulk dominates the surface) respectively. This is connected with a mesoscale regime where the innermost shell displays collective rotation and for outer shells the particles behave individually. 
 
Another interesting aspect of our study relates to the formation of pentagonal structures. Two-dimensional (2D) crystalline structures composed of pentagonal building blocks have garnered significant attention only in recent years due to their fundamental importance and possible uses in condensed matter physics \cite{caspar1996five}, material sciences \cite{zhuang2019pentagonal}, and mathematics \cite{ding2000tiling}. Regular pentagons are distinctive, in contrast to hexagons and squares, they do not encircle the entire plane without leaving spaces \cite{marcone2023polymorphous}. However, a total of $15$ irregular, convex pentagons were observed that can pave a plane completely without any gaps while yet maintaining periodicity \cite{zhuang2019pentagonal}. Pentagonal confinements are also employed in fluids to understand the behavior and characteristics of fluid convection patterns \cite{cerisier1991experimental}. 

While two-dimensional Yukawa dust clusters with hexagonal and square lattice structures have been talked about in the literature \cite{maity2019molecular}, we demonstrate here the formation of pentagonal structures. In fact, it appears whenever a two-dimensional dust cluster is confined in the plane and a new shell emerges in the arrangement with an increasing number of dust particles. This has been observed quite consistently.  It has been observed for a variety of confinement geometries and for different screening parameters. The pentagon structures are not only formed in dust clusters but also in electron-ion plasma in the ultracold regime. When an external perturbation like an oscillatory charge is introduced at the center of the cluster, it acts like a Langmuir probe. Due to this external perturbation, the pentagon structures are observed under certain conditions in the ultracold regime \cite{YADAV2024134326}.

 \begin{center}
    \textbf{ACKNOWLEDGEMENTS} 
 \end{center}
AD acknowledges support from the Anusandhan National Research Foundation (ANRF) of the Government of India through core grant  CRG/2022/002782 and J C Bose Fellowship grant JCB/2017/000055. The authors thank the IIT Delhi HPC facility for computational resources. M.Y. is thankful to the University Grants Commission (Grant No. 1316/CSIR-UGC NET DEC.2017) for funding the research.


\bibliography{dynamics_ref}

\begin{thebibliography}{48}%
\makeatletter
\providecommand \@ifxundefined [1]{%
 \@ifx{#1\undefined}
}%
\providecommand \@ifnum [1]{%
 \ifnum #1\expandafter \@firstoftwo
 \else \expandafter \@secondoftwo
 \fi
}%
\providecommand \@ifx [1]{%
 \ifx #1\expandafter \@firstoftwo
 \else \expandafter \@secondoftwo
 \fi
}%
\providecommand \natexlab [1]{#1}%
\providecommand \enquote  [1]{``#1''}%
\providecommand \bibnamefont  [1]{#1}%
\providecommand \bibfnamefont [1]{#1}%
\providecommand \citenamefont [1]{#1}%
\providecommand \href@noop [0]{\@secondoftwo}%
\providecommand \href [0]{\begingroup \@sanitize@url \@href}%
\providecommand \@href[1]{\@@startlink{#1}\@@href}%
\providecommand \@@href[1]{\endgroup#1\@@endlink}%
\providecommand \@sanitize@url [0]{\catcode `\\12\catcode `\$12\catcode
  `\&12\catcode `\#12\catcode `\^12\catcode `\_12\catcode `\%12\relax}%
\providecommand \@@startlink[1]{}%
\providecommand \@@endlink[0]{}%
\providecommand \url  [0]{\begingroup\@sanitize@url \@url }%
\providecommand \@url [1]{\endgroup\@href {#1}{\urlprefix }}%
\providecommand \urlprefix  [0]{URL }%
\providecommand \Eprint [0]{\href }%
\providecommand \doibase [0]{http://dx.doi.org/}%
\providecommand \selectlanguage [0]{\@gobble}%
\providecommand \bibinfo  [0]{\@secondoftwo}%
\providecommand \bibfield  [0]{\@secondoftwo}%
\providecommand \translation [1]{[#1]}%
\providecommand \BibitemOpen [0]{}%
\providecommand \bibitemStop [0]{}%
\providecommand \bibitemNoStop [0]{.\EOS\space}%
\providecommand \EOS [0]{\spacefactor3000\relax}%
\providecommand \BibitemShut  [1]{\csname bibitem#1\endcsname}%
\let\auto@bib@innerbib\@empty
\bibitem [{\citenamefont {Morfill}\ and\ \citenamefont
  {Ivlev}(2009)}]{morfill2009complex}%
  \BibitemOpen
  \bibfield  {author} {\bibinfo {author} {\bibfnamefont {G.~E.}\ \bibnamefont
  {Morfill}}\ and\ \bibinfo {author} {\bibfnamefont {A.~V.}\ \bibnamefont
  {Ivlev}},\ }\href@noop {} {\bibfield  {journal} {\bibinfo  {journal} {Reviews
  of modern physics}\ }\textbf {\bibinfo {volume} {81}},\ \bibinfo {pages}
  {1353} (\bibinfo {year} {2009})}\BibitemShut {NoStop}%
\bibitem [{\citenamefont {Hartmann}\ \emph {et~al.}(2007)\citenamefont
  {Hartmann}, \citenamefont {Donko}, \citenamefont {Bakshi}, \citenamefont
  {Kalman},\ and\ \citenamefont {Kyrkos}}]{hartmann2007molecular}%
  \BibitemOpen
  \bibfield  {author} {\bibinfo {author} {\bibfnamefont {P.}~\bibnamefont
  {Hartmann}}, \bibinfo {author} {\bibfnamefont {Z.}~\bibnamefont {Donko}},
  \bibinfo {author} {\bibfnamefont {P.~M.}\ \bibnamefont {Bakshi}}, \bibinfo
  {author} {\bibfnamefont {G.~J.}\ \bibnamefont {Kalman}}, \ and\ \bibinfo
  {author} {\bibfnamefont {S.}~\bibnamefont {Kyrkos}},\ }\href@noop {}
  {\bibfield  {journal} {\bibinfo  {journal} {IEEE transactions on plasma
  science}\ }\textbf {\bibinfo {volume} {35}},\ \bibinfo {pages} {332}
  (\bibinfo {year} {2007})}\BibitemShut {NoStop}%
\bibitem [{\citenamefont {Thomas}\ \emph {et~al.}(1994)\citenamefont {Thomas},
  \citenamefont {Morfill}, \citenamefont {Demmel}, \citenamefont {Goree},
  \citenamefont {Feuerbacher},\ and\ \citenamefont
  {M{\"o}hlmann}}]{thomas1994plasma}%
  \BibitemOpen
  \bibfield  {author} {\bibinfo {author} {\bibfnamefont {H.}~\bibnamefont
  {Thomas}}, \bibinfo {author} {\bibfnamefont {G.}~\bibnamefont {Morfill}},
  \bibinfo {author} {\bibfnamefont {V.}~\bibnamefont {Demmel}}, \bibinfo
  {author} {\bibfnamefont {J.}~\bibnamefont {Goree}}, \bibinfo {author}
  {\bibfnamefont {B.}~\bibnamefont {Feuerbacher}}, \ and\ \bibinfo {author}
  {\bibfnamefont {D.}~\bibnamefont {M{\"o}hlmann}},\ }\href@noop {} {\bibfield
  {journal} {\bibinfo  {journal} {Physical Review Letters}\ }\textbf {\bibinfo
  {volume} {73}},\ \bibinfo {pages} {652} (\bibinfo {year} {1994})}\BibitemShut
  {NoStop}%
\bibitem [{\citenamefont {Hayashi}\ and\ \citenamefont
  {Tachibana}(1994)}]{hayashi1994observation}%
  \BibitemOpen
  \bibfield  {author} {\bibinfo {author} {\bibfnamefont {Y.}~\bibnamefont
  {Hayashi}}\ and\ \bibinfo {author} {\bibfnamefont {K.}~\bibnamefont
  {Tachibana}},\ }\href@noop {} {\bibfield  {journal} {\bibinfo  {journal}
  {Jpn. J. Appl. Phys., Part}\ }\textbf {\bibinfo {volume} {2}},\ \bibinfo
  {pages} {L804} (\bibinfo {year} {1994})}\BibitemShut {NoStop}%
\bibitem [{\citenamefont {Chu}\ and\ \citenamefont
  {Lin}(1994)}]{chu1994direct}%
  \BibitemOpen
  \bibfield  {author} {\bibinfo {author} {\bibfnamefont {J.}~\bibnamefont
  {Chu}}\ and\ \bibinfo {author} {\bibfnamefont {I.}~\bibnamefont {Lin}},\
  }\href@noop {} {\bibfield  {journal} {\bibinfo  {journal} {Physical review
  letters}\ }\textbf {\bibinfo {volume} {72}},\ \bibinfo {pages} {4009}
  (\bibinfo {year} {1994})}\BibitemShut {NoStop}%
\bibitem [{\citenamefont {Feng}\ \emph {et~al.}(2010)\citenamefont {Feng},
  \citenamefont {Goree},\ and\ \citenamefont {Liu}}]{feng2010viscoelasticity}%
  \BibitemOpen
  \bibfield  {author} {\bibinfo {author} {\bibfnamefont {Y.}~\bibnamefont
  {Feng}}, \bibinfo {author} {\bibfnamefont {J.}~\bibnamefont {Goree}}, \ and\
  \bibinfo {author} {\bibfnamefont {B.}~\bibnamefont {Liu}},\ }\href@noop {}
  {\bibfield  {journal} {\bibinfo  {journal} {Physical review letters}\
  }\textbf {\bibinfo {volume} {105}},\ \bibinfo {pages} {025002} (\bibinfo
  {year} {2010})}\BibitemShut {NoStop}%
\bibitem [{\citenamefont {Feng}\ \emph {et~al.}(2012)\citenamefont {Feng},
  \citenamefont {Goree},\ and\ \citenamefont {Liu}}]{feng2012frequency}%
  \BibitemOpen
  \bibfield  {author} {\bibinfo {author} {\bibfnamefont {Y.}~\bibnamefont
  {Feng}}, \bibinfo {author} {\bibfnamefont {J.}~\bibnamefont {Goree}}, \ and\
  \bibinfo {author} {\bibfnamefont {B.}~\bibnamefont {Liu}},\ }\href@noop {}
  {\bibfield  {journal} {\bibinfo  {journal} {Physical Review E}\ }\textbf
  {\bibinfo {volume} {85}},\ \bibinfo {pages} {066402} (\bibinfo {year}
  {2012})}\BibitemShut {NoStop}%
\bibitem [{\citenamefont {Singh~Dharodi}\ \emph {et~al.}(2014)\citenamefont
  {Singh~Dharodi}, \citenamefont {Kumar~Tiwari},\ and\ \citenamefont
  {Das}}]{singh2014visco}%
  \BibitemOpen
  \bibfield  {author} {\bibinfo {author} {\bibfnamefont {V.}~\bibnamefont
  {Singh~Dharodi}}, \bibinfo {author} {\bibfnamefont {S.}~\bibnamefont
  {Kumar~Tiwari}}, \ and\ \bibinfo {author} {\bibfnamefont {A.}~\bibnamefont
  {Das}},\ }\href@noop {} {\bibfield  {journal} {\bibinfo  {journal} {Physics
  of Plasmas}\ }\textbf {\bibinfo {volume} {21}} (\bibinfo {year}
  {2014})}\BibitemShut {NoStop}%
\bibitem [{\citenamefont {Nosenko}\ \emph {et~al.}(2008)\citenamefont
  {Nosenko}, \citenamefont {Zhdanov}, \citenamefont {Ivlev}, \citenamefont
  {Morfill}, \citenamefont {Goree},\ and\ \citenamefont
  {Piel}}]{nosenko2008heat}%
  \BibitemOpen
  \bibfield  {author} {\bibinfo {author} {\bibfnamefont {V.}~\bibnamefont
  {Nosenko}}, \bibinfo {author} {\bibfnamefont {S.}~\bibnamefont {Zhdanov}},
  \bibinfo {author} {\bibfnamefont {A.}~\bibnamefont {Ivlev}}, \bibinfo
  {author} {\bibfnamefont {G.}~\bibnamefont {Morfill}}, \bibinfo {author}
  {\bibfnamefont {J.}~\bibnamefont {Goree}}, \ and\ \bibinfo {author}
  {\bibfnamefont {A.}~\bibnamefont {Piel}},\ }\href@noop {} {\bibfield
  {journal} {\bibinfo  {journal} {Physical review letters}\ }\textbf {\bibinfo
  {volume} {100}},\ \bibinfo {pages} {025003} (\bibinfo {year}
  {2008})}\BibitemShut {NoStop}%
\bibitem [{\citenamefont {Nunomura}\ \emph {et~al.}(2005)\citenamefont
  {Nunomura}, \citenamefont {Samsonov}, \citenamefont {Zhdanov},\ and\
  \citenamefont {Morfill}}]{nunomura2005heat}%
  \BibitemOpen
  \bibfield  {author} {\bibinfo {author} {\bibfnamefont {S.}~\bibnamefont
  {Nunomura}}, \bibinfo {author} {\bibfnamefont {D.}~\bibnamefont {Samsonov}},
  \bibinfo {author} {\bibfnamefont {S.}~\bibnamefont {Zhdanov}}, \ and\
  \bibinfo {author} {\bibfnamefont {G.}~\bibnamefont {Morfill}},\ }\href@noop
  {} {\bibfield  {journal} {\bibinfo  {journal} {Physical review letters}\
  }\textbf {\bibinfo {volume} {95}},\ \bibinfo {pages} {025003} (\bibinfo
  {year} {2005})}\BibitemShut {NoStop}%
\bibitem [{\citenamefont {Liu}\ and\ \citenamefont
  {Goree}(2008)}]{liu2008superdiffusion}%
  \BibitemOpen
  \bibfield  {author} {\bibinfo {author} {\bibfnamefont {B.}~\bibnamefont
  {Liu}}\ and\ \bibinfo {author} {\bibfnamefont {J.}~\bibnamefont {Goree}},\
  }\href@noop {} {\bibfield  {journal} {\bibinfo  {journal} {Physical review
  letters}\ }\textbf {\bibinfo {volume} {100}},\ \bibinfo {pages} {055003}
  (\bibinfo {year} {2008})}\BibitemShut {NoStop}%
\bibitem [{\citenamefont {Merlino}\ \emph {et~al.}(1998)\citenamefont
  {Merlino}, \citenamefont {Barkan}, \citenamefont {Thompson},\ and\
  \citenamefont {D’angelo}}]{merlino1998laboratory}%
  \BibitemOpen
  \bibfield  {author} {\bibinfo {author} {\bibfnamefont {R.}~\bibnamefont
  {Merlino}}, \bibinfo {author} {\bibfnamefont {A.}~\bibnamefont {Barkan}},
  \bibinfo {author} {\bibfnamefont {C.}~\bibnamefont {Thompson}}, \ and\
  \bibinfo {author} {\bibfnamefont {N.}~\bibnamefont {D’angelo}},\
  }\href@noop {} {\bibfield  {journal} {\bibinfo  {journal} {Physics of
  Plasmas}\ }\textbf {\bibinfo {volume} {5}},\ \bibinfo {pages} {1607}
  (\bibinfo {year} {1998})}\BibitemShut {NoStop}%
\bibitem [{\citenamefont {Tiwari}\ \emph {et~al.}(2012)\citenamefont {Tiwari},
  \citenamefont {Das}, \citenamefont {Angom}, \citenamefont {Patel},\ and\
  \citenamefont {Kaw}}]{tiwari2012kelvin}%
  \BibitemOpen
  \bibfield  {author} {\bibinfo {author} {\bibfnamefont {S.~K.}\ \bibnamefont
  {Tiwari}}, \bibinfo {author} {\bibfnamefont {A.}~\bibnamefont {Das}},
  \bibinfo {author} {\bibfnamefont {D.}~\bibnamefont {Angom}}, \bibinfo
  {author} {\bibfnamefont {B.~G.}\ \bibnamefont {Patel}}, \ and\ \bibinfo
  {author} {\bibfnamefont {P.}~\bibnamefont {Kaw}},\ }\href@noop {} {\bibfield
  {journal} {\bibinfo  {journal} {Physics of Plasmas}\ }\textbf {\bibinfo
  {volume} {19}} (\bibinfo {year} {2012})}\BibitemShut {NoStop}%
\bibitem [{\citenamefont {Das}\ and\ \citenamefont
  {Kaw}(2014)}]{das2014suppression}%
  \BibitemOpen
  \bibfield  {author} {\bibinfo {author} {\bibfnamefont {A.}~\bibnamefont
  {Das}}\ and\ \bibinfo {author} {\bibfnamefont {P.}~\bibnamefont {Kaw}},\
  }\href@noop {} {\bibfield  {journal} {\bibinfo  {journal} {Physics of
  Plasmas}\ }\textbf {\bibinfo {volume} {21}} (\bibinfo {year}
  {2014})}\BibitemShut {NoStop}%
\bibitem [{\citenamefont {Dharodi}\ \emph {et~al.}(2022)\citenamefont
  {Dharodi}, \citenamefont {Patel},\ and\ \citenamefont
  {Das}}]{dharodi2022kelvin}%
  \BibitemOpen
  \bibfield  {author} {\bibinfo {author} {\bibfnamefont {V.~S.}\ \bibnamefont
  {Dharodi}}, \bibinfo {author} {\bibfnamefont {B.}~\bibnamefont {Patel}}, \
  and\ \bibinfo {author} {\bibfnamefont {A.}~\bibnamefont {Das}},\ }\href@noop
  {} {\bibfield  {journal} {\bibinfo  {journal} {Journal of Plasma Physics}\
  }\textbf {\bibinfo {volume} {88}},\ \bibinfo {pages} {905880103} (\bibinfo
  {year} {2022})}\BibitemShut {NoStop}%
\bibitem [{\citenamefont {Veeresha}\ \emph {et~al.}(2005)\citenamefont
  {Veeresha}, \citenamefont {Das},\ and\ \citenamefont
  {Sen}}]{veeresha2005rayleigh}%
  \BibitemOpen
  \bibfield  {author} {\bibinfo {author} {\bibfnamefont {B.}~\bibnamefont
  {Veeresha}}, \bibinfo {author} {\bibfnamefont {A.}~\bibnamefont {Das}}, \
  and\ \bibinfo {author} {\bibfnamefont {A.}~\bibnamefont {Sen}},\ }\href@noop
  {} {\bibfield  {journal} {\bibinfo  {journal} {Physics of plasmas}\ }\textbf
  {\bibinfo {volume} {12}} (\bibinfo {year} {2005})}\BibitemShut {NoStop}%
\bibitem [{\citenamefont {Kumar}\ \emph {et~al.}(2018)\citenamefont {Kumar},
  \citenamefont {Patel},\ and\ \citenamefont {Das}}]{kumar2018spiral}%
  \BibitemOpen
  \bibfield  {author} {\bibinfo {author} {\bibfnamefont {S.}~\bibnamefont
  {Kumar}}, \bibinfo {author} {\bibfnamefont {B.}~\bibnamefont {Patel}}, \ and\
  \bibinfo {author} {\bibfnamefont {A.}~\bibnamefont {Das}},\ }\href@noop {}
  {\bibfield  {journal} {\bibinfo  {journal} {Physics of Plasmas}\ }\textbf
  {\bibinfo {volume} {25}} (\bibinfo {year} {2018})}\BibitemShut {NoStop}%
\bibitem [{\citenamefont {Das}\ \emph {et~al.}(2014)\citenamefont {Das},
  \citenamefont {Tiwari}, \citenamefont {Kaw},\ and\ \citenamefont
  {Sen}}]{das2014exact}%
  \BibitemOpen
  \bibfield  {author} {\bibinfo {author} {\bibfnamefont {A.}~\bibnamefont
  {Das}}, \bibinfo {author} {\bibfnamefont {S.~K.}\ \bibnamefont {Tiwari}},
  \bibinfo {author} {\bibfnamefont {P.}~\bibnamefont {Kaw}}, \ and\ \bibinfo
  {author} {\bibfnamefont {A.}~\bibnamefont {Sen}},\ }\href@noop {} {\bibfield
  {journal} {\bibinfo  {journal} {Physics of Plasmas}\ }\textbf {\bibinfo
  {volume} {21}} (\bibinfo {year} {2014})}\BibitemShut {NoStop}%
\bibitem [{\citenamefont {Kumar}\ \emph {et~al.}(2017)\citenamefont {Kumar},
  \citenamefont {Tiwari},\ and\ \citenamefont {Das}}]{kumar2017observation}%
  \BibitemOpen
  \bibfield  {author} {\bibinfo {author} {\bibfnamefont {S.}~\bibnamefont
  {Kumar}}, \bibinfo {author} {\bibfnamefont {S.~K.}\ \bibnamefont {Tiwari}}, \
  and\ \bibinfo {author} {\bibfnamefont {A.}~\bibnamefont {Das}},\ }\href@noop
  {} {\bibfield  {journal} {\bibinfo  {journal} {Physics of Plasmas}\ }\textbf
  {\bibinfo {volume} {24}} (\bibinfo {year} {2017})}\BibitemShut {NoStop}%
\bibitem [{\citenamefont {Kaw}\ and\ \citenamefont {Sen}(1998)}]{kaw1998low}%
  \BibitemOpen
  \bibfield  {author} {\bibinfo {author} {\bibfnamefont {P.}~\bibnamefont
  {Kaw}}\ and\ \bibinfo {author} {\bibfnamefont {A.}~\bibnamefont {Sen}},\
  }\href@noop {} {\bibfield  {journal} {\bibinfo  {journal} {Physics of
  Plasmas}\ }\textbf {\bibinfo {volume} {5}},\ \bibinfo {pages} {3552}
  (\bibinfo {year} {1998})}\BibitemShut {NoStop}%
\bibitem [{\citenamefont {Kaw}(2001)}]{kaw2001collective}%
  \BibitemOpen
  \bibfield  {author} {\bibinfo {author} {\bibfnamefont {P.}~\bibnamefont
  {Kaw}},\ }\href@noop {} {\bibfield  {journal} {\bibinfo  {journal} {Physics
  of Plasmas}\ }\textbf {\bibinfo {volume} {8}},\ \bibinfo {pages} {1870}
  (\bibinfo {year} {2001})}\BibitemShut {NoStop}%
\bibitem [{\citenamefont {Ivlev}\ and\ \citenamefont
  {Morfill}(2000)}]{ivlev2000anisotropic}%
  \BibitemOpen
  \bibfield  {author} {\bibinfo {author} {\bibfnamefont {A.}~\bibnamefont
  {Ivlev}}\ and\ \bibinfo {author} {\bibfnamefont {G.}~\bibnamefont
  {Morfill}},\ }\href@noop {} {\bibfield  {journal} {\bibinfo  {journal}
  {Physical Review E}\ }\textbf {\bibinfo {volume} {63}},\ \bibinfo {pages}
  {016409} (\bibinfo {year} {2000})}\BibitemShut {NoStop}%
\bibitem [{\citenamefont {Thomas}\ and\ \citenamefont
  {Morfill}(1996)}]{thomas1996melting}%
  \BibitemOpen
  \bibfield  {author} {\bibinfo {author} {\bibfnamefont {H.~M.}\ \bibnamefont
  {Thomas}}\ and\ \bibinfo {author} {\bibfnamefont {G.~E.}\ \bibnamefont
  {Morfill}},\ }\href@noop {} {\bibfield  {journal} {\bibinfo  {journal}
  {Nature}\ }\textbf {\bibinfo {volume} {379}},\ \bibinfo {pages} {806}
  (\bibinfo {year} {1996})}\BibitemShut {NoStop}%
\bibitem [{\citenamefont {Melzer}\ \emph
  {et~al.}(1996{\natexlab{a}})\citenamefont {Melzer}, \citenamefont {Homann},\
  and\ \citenamefont {Piel}}]{melzer1996experimental}%
  \BibitemOpen
  \bibfield  {author} {\bibinfo {author} {\bibfnamefont {A.}~\bibnamefont
  {Melzer}}, \bibinfo {author} {\bibfnamefont {A.}~\bibnamefont {Homann}}, \
  and\ \bibinfo {author} {\bibfnamefont {A.}~\bibnamefont {Piel}},\ }\href@noop
  {} {\bibfield  {journal} {\bibinfo  {journal} {Physical Review E}\ }\textbf
  {\bibinfo {volume} {53}},\ \bibinfo {pages} {2757} (\bibinfo {year}
  {1996}{\natexlab{a}})}\BibitemShut {NoStop}%
\bibitem [{\citenamefont {Melzer}\ \emph
  {et~al.}(1996{\natexlab{b}})\citenamefont {Melzer}, \citenamefont
  {Schweigert}, \citenamefont {Schweigert}, \citenamefont {Homann},
  \citenamefont {Peters},\ and\ \citenamefont {Piel}}]{melzer1996structure}%
  \BibitemOpen
  \bibfield  {author} {\bibinfo {author} {\bibfnamefont {A.}~\bibnamefont
  {Melzer}}, \bibinfo {author} {\bibfnamefont {V.}~\bibnamefont {Schweigert}},
  \bibinfo {author} {\bibfnamefont {I.}~\bibnamefont {Schweigert}}, \bibinfo
  {author} {\bibfnamefont {A.}~\bibnamefont {Homann}}, \bibinfo {author}
  {\bibfnamefont {S.}~\bibnamefont {Peters}}, \ and\ \bibinfo {author}
  {\bibfnamefont {A.}~\bibnamefont {Piel}},\ }\href@noop {} {\bibfield
  {journal} {\bibinfo  {journal} {Physical Review E}\ }\textbf {\bibinfo
  {volume} {54}},\ \bibinfo {pages} {R46} (\bibinfo {year}
  {1996}{\natexlab{b}})}\BibitemShut {NoStop}%
\bibitem [{\citenamefont {Maity}\ and\ \citenamefont
  {Das}(2019)}]{maity2019molecular}%
  \BibitemOpen
  \bibfield  {author} {\bibinfo {author} {\bibfnamefont {S.}~\bibnamefont
  {Maity}}\ and\ \bibinfo {author} {\bibfnamefont {A.}~\bibnamefont {Das}},\
  }\href@noop {} {\bibfield  {journal} {\bibinfo  {journal} {Physics of
  Plasmas}\ }\textbf {\bibinfo {volume} {26}} (\bibinfo {year}
  {2019})}\BibitemShut {NoStop}%
\bibitem [{\citenamefont {Schweigert}\ \emph {et~al.}(1998)\citenamefont
  {Schweigert}, \citenamefont {Schweigert}, \citenamefont {Melzer},
  \citenamefont {Homann},\ and\ \citenamefont {Piel}}]{schweigert1998plasma}%
  \BibitemOpen
  \bibfield  {author} {\bibinfo {author} {\bibfnamefont {V.}~\bibnamefont
  {Schweigert}}, \bibinfo {author} {\bibfnamefont {I.}~\bibnamefont
  {Schweigert}}, \bibinfo {author} {\bibfnamefont {A.}~\bibnamefont {Melzer}},
  \bibinfo {author} {\bibfnamefont {A.}~\bibnamefont {Homann}}, \ and\ \bibinfo
  {author} {\bibfnamefont {A.}~\bibnamefont {Piel}},\ }\href@noop {} {\bibfield
   {journal} {\bibinfo  {journal} {Physical review letters}\ }\textbf {\bibinfo
  {volume} {80}},\ \bibinfo {pages} {5345} (\bibinfo {year}
  {1998})}\BibitemShut {NoStop}%
\bibitem [{\citenamefont {Teng}\ \emph {et~al.}(2009)\citenamefont {Teng},
  \citenamefont {Chang}, \citenamefont {Tseng},\ and\ \citenamefont
  {Lin}}]{teng2009wave}%
  \BibitemOpen
  \bibfield  {author} {\bibinfo {author} {\bibfnamefont {L.-W.}\ \bibnamefont
  {Teng}}, \bibinfo {author} {\bibfnamefont {M.-C.}\ \bibnamefont {Chang}},
  \bibinfo {author} {\bibfnamefont {Y.-P.}\ \bibnamefont {Tseng}}, \ and\
  \bibinfo {author} {\bibfnamefont {I.}~\bibnamefont {Lin}},\ }\href@noop {}
  {\bibfield  {journal} {\bibinfo  {journal} {Physical review letters}\
  }\textbf {\bibinfo {volume} {103}},\ \bibinfo {pages} {245005} (\bibinfo
  {year} {2009})}\BibitemShut {NoStop}%
\bibitem [{\citenamefont {Maity}\ \emph {et~al.}(2018)\citenamefont {Maity},
  \citenamefont {Das}, \citenamefont {Kumar},\ and\ \citenamefont
  {Tiwari}}]{maity2018interplay}%
  \BibitemOpen
  \bibfield  {author} {\bibinfo {author} {\bibfnamefont {S.}~\bibnamefont
  {Maity}}, \bibinfo {author} {\bibfnamefont {A.}~\bibnamefont {Das}}, \bibinfo
  {author} {\bibfnamefont {S.}~\bibnamefont {Kumar}}, \ and\ \bibinfo {author}
  {\bibfnamefont {S.~K.}\ \bibnamefont {Tiwari}},\ }\href@noop {} {\bibfield
  {journal} {\bibinfo  {journal} {Physics of Plasmas}\ }\textbf {\bibinfo
  {volume} {25}} (\bibinfo {year} {2018})}\BibitemShut {NoStop}%
\bibitem [{\citenamefont {Maity}\ \emph {et~al.}(2020)\citenamefont {Maity},
  \citenamefont {Deshwal}, \citenamefont {Yadav},\ and\ \citenamefont
  {Das}}]{maity2020dynamical}%
  \BibitemOpen
  \bibfield  {author} {\bibinfo {author} {\bibfnamefont {S.}~\bibnamefont
  {Maity}}, \bibinfo {author} {\bibfnamefont {P.}~\bibnamefont {Deshwal}},
  \bibinfo {author} {\bibfnamefont {M.}~\bibnamefont {Yadav}}, \ and\ \bibinfo
  {author} {\bibfnamefont {A.}~\bibnamefont {Das}},\ }\href@noop {} {\bibfield
  {journal} {\bibinfo  {journal} {Physical Review E}\ }\textbf {\bibinfo
  {volume} {102}},\ \bibinfo {pages} {023213} (\bibinfo {year}
  {2020})}\BibitemShut {NoStop}%
\bibitem [{\citenamefont {Deshwal}\ \emph {et~al.}(2022)\citenamefont
  {Deshwal}, \citenamefont {Yadav}, \citenamefont {Prasad}, \citenamefont
  {Sridev}, \citenamefont {Ahuja}, \citenamefont {Maity},\ and\ \citenamefont
  {Das}}]{deshwal2022chaotic}%
  \BibitemOpen
  \bibfield  {author} {\bibinfo {author} {\bibfnamefont {P.}~\bibnamefont
  {Deshwal}}, \bibinfo {author} {\bibfnamefont {M.}~\bibnamefont {Yadav}},
  \bibinfo {author} {\bibfnamefont {C.}~\bibnamefont {Prasad}}, \bibinfo
  {author} {\bibfnamefont {S.}~\bibnamefont {Sridev}}, \bibinfo {author}
  {\bibfnamefont {Y.}~\bibnamefont {Ahuja}}, \bibinfo {author} {\bibfnamefont
  {S.}~\bibnamefont {Maity}}, \ and\ \bibinfo {author} {\bibfnamefont
  {A.}~\bibnamefont {Das}},\ }\href@noop {} {\bibfield  {journal} {\bibinfo
  {journal} {Chaos: An Interdisciplinary Journal of Nonlinear Science}\
  }\textbf {\bibinfo {volume} {32}} (\bibinfo {year} {2022})}\BibitemShut
  {NoStop}%
\bibitem [{\citenamefont {Robicheaux}\ and\ \citenamefont
  {Hanson}(2002)}]{robicheaux2002simulation}%
  \BibitemOpen
  \bibfield  {author} {\bibinfo {author} {\bibfnamefont {F.}~\bibnamefont
  {Robicheaux}}\ and\ \bibinfo {author} {\bibfnamefont {J.~D.}\ \bibnamefont
  {Hanson}},\ }\href@noop {} {\bibfield  {journal} {\bibinfo  {journal}
  {Physical review letters}\ }\textbf {\bibinfo {volume} {88}},\ \bibinfo
  {pages} {055002} (\bibinfo {year} {2002})}\BibitemShut {NoStop}%
\bibitem [{\citenamefont {Pohl}\ \emph {et~al.}(2004)\citenamefont {Pohl},
  \citenamefont {Pattard},\ and\ \citenamefont {Rost}}]{pohl2004kinetic}%
  \BibitemOpen
  \bibfield  {author} {\bibinfo {author} {\bibfnamefont {T.}~\bibnamefont
  {Pohl}}, \bibinfo {author} {\bibfnamefont {T.}~\bibnamefont {Pattard}}, \
  and\ \bibinfo {author} {\bibfnamefont {J.~M.}\ \bibnamefont {Rost}},\
  }\href@noop {} {\bibfield  {journal} {\bibinfo  {journal} {Physical Review
  A}\ }\textbf {\bibinfo {volume} {70}},\ \bibinfo {pages} {033416} (\bibinfo
  {year} {2004})}\BibitemShut {NoStop}%
\bibitem [{\citenamefont {Laha}\ \emph {et~al.}(2007)\citenamefont {Laha},
  \citenamefont {Gupta}, \citenamefont {Simien}, \citenamefont {Gao},
  \citenamefont {Castro}, \citenamefont {Pohl},\ and\ \citenamefont
  {Killian}}]{laha2007experimental}%
  \BibitemOpen
  \bibfield  {author} {\bibinfo {author} {\bibfnamefont {S.}~\bibnamefont
  {Laha}}, \bibinfo {author} {\bibfnamefont {P.}~\bibnamefont {Gupta}},
  \bibinfo {author} {\bibfnamefont {C.}~\bibnamefont {Simien}}, \bibinfo
  {author} {\bibfnamefont {H.}~\bibnamefont {Gao}}, \bibinfo {author}
  {\bibfnamefont {J.}~\bibnamefont {Castro}}, \bibinfo {author} {\bibfnamefont
  {T.}~\bibnamefont {Pohl}}, \ and\ \bibinfo {author} {\bibfnamefont
  {T.}~\bibnamefont {Killian}},\ }\href@noop {} {\bibfield  {journal} {\bibinfo
   {journal} {Physical review letters}\ }\textbf {\bibinfo {volume} {99}},\
  \bibinfo {pages} {155001} (\bibinfo {year} {2007})}\BibitemShut {NoStop}%
\bibitem [{\citenamefont {Killian}\ \emph {et~al.}(2001)\citenamefont
  {Killian}, \citenamefont {Lim}, \citenamefont {Kulin}, \citenamefont {Dumke},
  \citenamefont {Bergeson},\ and\ \citenamefont
  {Rolston}}]{killian2001formation}%
  \BibitemOpen
  \bibfield  {author} {\bibinfo {author} {\bibfnamefont {T.}~\bibnamefont
  {Killian}}, \bibinfo {author} {\bibfnamefont {M.}~\bibnamefont {Lim}},
  \bibinfo {author} {\bibfnamefont {S.}~\bibnamefont {Kulin}}, \bibinfo
  {author} {\bibfnamefont {R.}~\bibnamefont {Dumke}}, \bibinfo {author}
  {\bibfnamefont {S.}~\bibnamefont {Bergeson}}, \ and\ \bibinfo {author}
  {\bibfnamefont {S.}~\bibnamefont {Rolston}},\ }\href@noop {} {\bibfield
  {journal} {\bibinfo  {journal} {Physical review letters}\ }\textbf {\bibinfo
  {volume} {86}},\ \bibinfo {pages} {3759} (\bibinfo {year}
  {2001})}\BibitemShut {NoStop}%
\bibitem [{\citenamefont {Yadav}\ \emph {et~al.}(2023)\citenamefont {Yadav},
  \citenamefont {Deshwal}, \citenamefont {Maity},\ and\ \citenamefont
  {Das}}]{yadav2023structure}%
  \BibitemOpen
  \bibfield  {author} {\bibinfo {author} {\bibfnamefont {M.}~\bibnamefont
  {Yadav}}, \bibinfo {author} {\bibfnamefont {P.}~\bibnamefont {Deshwal}},
  \bibinfo {author} {\bibfnamefont {S.}~\bibnamefont {Maity}}, \ and\ \bibinfo
  {author} {\bibfnamefont {A.}~\bibnamefont {Das}},\ }\href@noop {} {\bibfield
  {journal} {\bibinfo  {journal} {Physical Review E}\ }\textbf {\bibinfo
  {volume} {107}},\ \bibinfo {pages} {055214} (\bibinfo {year}
  {2023})}\BibitemShut {NoStop}%
\bibitem [{\citenamefont {Yadav}\ \emph {et~al.}(2024)\citenamefont {Yadav},
  \citenamefont {Katariya}, \citenamefont {Sharma},\ and\ \citenamefont
  {Das}}]{YADAV2024134326}%
  \BibitemOpen
  \bibfield  {author} {\bibinfo {author} {\bibfnamefont {M.}~\bibnamefont
  {Yadav}}, \bibinfo {author} {\bibfnamefont {A.~S.}\ \bibnamefont {Katariya}},
  \bibinfo {author} {\bibfnamefont {A.}~\bibnamefont {Sharma}}, \ and\ \bibinfo
  {author} {\bibfnamefont {A.}~\bibnamefont {Das}},\ }\href {\doibase
  https://doi.org/10.1016/j.physd.2024.134326} {\bibfield  {journal} {\bibinfo
  {journal} {Physica D: Nonlinear Phenomena}\ }\textbf {\bibinfo {volume}
  {469}},\ \bibinfo {pages} {134326} (\bibinfo {year} {2024})}\BibitemShut
  {NoStop}%
\bibitem [{\citenamefont {Van~Horn}(1991)}]{van1991dense}%
  \BibitemOpen
  \bibfield  {author} {\bibinfo {author} {\bibfnamefont {H.}~\bibnamefont
  {Van~Horn}},\ }\href@noop {} {\bibfield  {journal} {\bibinfo  {journal}
  {Science}\ }\textbf {\bibinfo {volume} {252}},\ \bibinfo {pages} {384}
  (\bibinfo {year} {1991})}\BibitemShut {NoStop}%
\bibitem [{\citenamefont {Plimpton}(1995)}]{plimpton1995fast}%
  \BibitemOpen
  \bibfield  {author} {\bibinfo {author} {\bibfnamefont {S.}~\bibnamefont
  {Plimpton}},\ }\href@noop {} {\bibfield  {journal} {\bibinfo  {journal}
  {Journal of computational physics}\ }\textbf {\bibinfo {volume} {117}},\
  \bibinfo {pages} {1} (\bibinfo {year} {1995})}\BibitemShut {NoStop}%
\bibitem [{\citenamefont {Nosenko}\ and\ \citenamefont
  {Goree}(2004)}]{nosenko2004shear}%
  \BibitemOpen
  \bibfield  {author} {\bibinfo {author} {\bibfnamefont {V.}~\bibnamefont
  {Nosenko}}\ and\ \bibinfo {author} {\bibfnamefont {J.}~\bibnamefont
  {Goree}},\ }\href@noop {} {\bibfield  {journal} {\bibinfo  {journal}
  {Physical review letters}\ }\textbf {\bibinfo {volume} {93}},\ \bibinfo
  {pages} {155004} (\bibinfo {year} {2004})}\BibitemShut {NoStop}%
\bibitem [{\citenamefont {Shukla}\ and\ \citenamefont
  {Mamun}(2015)}]{shukla2015introduction}%
  \BibitemOpen
  \bibfield  {author} {\bibinfo {author} {\bibfnamefont {P.~K.}\ \bibnamefont
  {Shukla}}\ and\ \bibinfo {author} {\bibfnamefont {A.}~\bibnamefont {Mamun}},\
  }\href@noop {} {\emph {\bibinfo {title} {Introduction to dusty plasma
  physics}}}\ (\bibinfo  {publisher} {CRC press},\ \bibinfo {year}
  {2015})\BibitemShut {NoStop}%
\bibitem [{\citenamefont {Nos{\'e}}(1984)}]{nose1984molecular}%
  \BibitemOpen
  \bibfield  {author} {\bibinfo {author} {\bibfnamefont {S.}~\bibnamefont
  {Nos{\'e}}},\ }\href@noop {} {\bibfield  {journal} {\bibinfo  {journal}
  {Molecular physics}\ }\textbf {\bibinfo {volume} {52}},\ \bibinfo {pages}
  {255} (\bibinfo {year} {1984})}\BibitemShut {NoStop}%
\bibitem [{\citenamefont {Hoover}(1985)}]{PhysRevA.31.1695}%
  \BibitemOpen
  \bibfield  {author} {\bibinfo {author} {\bibfnamefont {W.~G.}\ \bibnamefont
  {Hoover}},\ }\href {\doibase 10.1103/PhysRevA.31.1695} {\bibfield  {journal}
  {\bibinfo  {journal} {Phys. Rev. A}\ }\textbf {\bibinfo {volume} {31}},\
  \bibinfo {pages} {1695} (\bibinfo {year} {1985})}\BibitemShut {NoStop}%
\bibitem [{\citenamefont {Caspar}\ and\ \citenamefont
  {Fontano}(1996)}]{caspar1996five}%
  \BibitemOpen
  \bibfield  {author} {\bibinfo {author} {\bibfnamefont {D.~L.}\ \bibnamefont
  {Caspar}}\ and\ \bibinfo {author} {\bibfnamefont {E.}~\bibnamefont
  {Fontano}},\ }\href@noop {} {\bibfield  {journal} {\bibinfo  {journal}
  {Proceedings of the National Academy of Sciences}\ }\textbf {\bibinfo
  {volume} {93}},\ \bibinfo {pages} {14271} (\bibinfo {year}
  {1996})}\BibitemShut {NoStop}%
\bibitem [{\citenamefont {Zhuang}(2019)}]{zhuang2019pentagonal}%
  \BibitemOpen
  \bibfield  {author} {\bibinfo {author} {\bibfnamefont {H.~L.}\ \bibnamefont
  {Zhuang}},\ }\href@noop {} {\bibfield  {journal} {\bibinfo  {journal}
  {Computational Materials Science}\ }\textbf {\bibinfo {volume} {159}},\
  \bibinfo {pages} {448} (\bibinfo {year} {2019})}\BibitemShut {NoStop}%
\bibitem [{\citenamefont {Ding}\ \emph {et~al.}(2000)\citenamefont {Ding},
  \citenamefont {Schattschneider},\ and\ \citenamefont
  {Zamfirescu}}]{ding2000tiling}%
  \BibitemOpen
  \bibfield  {author} {\bibinfo {author} {\bibfnamefont {R.}~\bibnamefont
  {Ding}}, \bibinfo {author} {\bibfnamefont {D.}~\bibnamefont
  {Schattschneider}}, \ and\ \bibinfo {author} {\bibfnamefont {T.}~\bibnamefont
  {Zamfirescu}},\ }\href@noop {} {\bibfield  {journal} {\bibinfo  {journal}
  {Discrete Mathematics}\ }\textbf {\bibinfo {volume} {221}},\ \bibinfo {pages}
  {113} (\bibinfo {year} {2000})}\BibitemShut {NoStop}%
\bibitem [{\citenamefont {Marcone}\ \emph {et~al.}(2023)\citenamefont
  {Marcone}, \citenamefont {Cha{\^a}bani}, \citenamefont {Goldmann},
  \citenamefont {Imp{\'e}ror-Clerc}, \citenamefont {Constantin},\ and\
  \citenamefont {Hamon}}]{marcone2023polymorphous}%
  \BibitemOpen
  \bibfield  {author} {\bibinfo {author} {\bibfnamefont {J.}~\bibnamefont
  {Marcone}}, \bibinfo {author} {\bibfnamefont {W.}~\bibnamefont
  {Cha{\^a}bani}}, \bibinfo {author} {\bibfnamefont {C.}~\bibnamefont
  {Goldmann}}, \bibinfo {author} {\bibfnamefont {M.}~\bibnamefont
  {Imp{\'e}ror-Clerc}}, \bibinfo {author} {\bibfnamefont {D.}~\bibnamefont
  {Constantin}}, \ and\ \bibinfo {author} {\bibfnamefont {C.}~\bibnamefont
  {Hamon}},\ }\href@noop {} {\bibfield  {journal} {\bibinfo  {journal} {Nano
  Letters}\ }\textbf {\bibinfo {volume} {23}},\ \bibinfo {pages} {1337}
  (\bibinfo {year} {2023})}\BibitemShut {NoStop}%
\bibitem [{\citenamefont {Cerisier}(1991)}]{cerisier1991experimental}%
  \BibitemOpen
  \bibfield  {author} {\bibinfo {author} {\bibfnamefont {P.}~\bibnamefont
  {Cerisier}},\ }\href@noop {} {\bibfield  {journal} {\bibinfo  {journal}
  {Physics of Fluids A: Fluid Dynamics}\ }\textbf {\bibinfo {volume} {3}},\
  \bibinfo {pages} {2061} (\bibinfo {year} {1991})}\BibitemShut {NoStop}%
\end{thebibliography}%
\end{document}